\documentclass[]{aa} 
\usepackage{graphicx}
\usepackage{natbib}
\usepackage{txfonts}
\begin{document}
   \title{A Catalog of Extended Clusters and Ultra-Compact Dwarf Galaxies}

\titlerunning{A Catalog of ECs and UCDs}

   \subtitle{An Analysis of their Parameters in Early- and Late-Type Galaxies}

\author{R.C. Br\"uns\inst{1} \and P. Kroupa\inst{1}}

\institute{Argelander-Institut f\"ur Astronomie, Universit\"at Bonn,
            Auf dem H\"ugel 71, D-53121 Bonn, Germany\\
              \email{rcbruens@astro.uni-bonn.de, pavel@astro.uni-bonn.de}
         }
	     
   \date{Received 28 May 2012; accepted 14 September 2012}

 
  \abstract
   {In the last decade, very extended old stellar clusters with masses in the range from
  a few $10^{4}$ to $10^{8}$ M$_{\sun}$ and effective radii larger than 10 pc have been found in 
  various types of galaxies in different environments. Objects with masses comparable
to normal globular clusters (GCs) are called extended clusters (ECs), while objects with
masses in the dwarf galaxy regime are called ultra-compact dwarf galaxies (UCDs).}
   {The paper analyses the observational parameters total luminosity, $M_{\rm V}$, effective radius, $r_{\rm eff}$, 
   and projected distance to the host galaxy, $R_{\rm proj}$, of all known ECs and UCDs and the dependence 
   of these parameters on the type and the total luminosity of their host galaxy.}
   {We searched the available literature to compile a catalog of star clusters 
   with effective radii larger than 10 pc. As there is no clear distinction between ECs and UCDs, 
   both types of objects will be called extended stellar objects -- abbreviated "EOs" -- in this paper.}
   {In total, we found 813 EOs of which 171 are associated with late-type galaxies and 642 EOs associated 
   with early-type galaxies. EOs cover a luminosity range from about $M_{\rm V} = -4$ to $-$14 mag. 
   However, the vast majority of EOs brighter than $M_{\rm V} = -10$ mag 
   are associated with giant elliptical galaxies. At each magnitude extended objects are found with 
   effective radii between 10 pc and an upper size limit, which shows a clear trend: the more 
   luminous the object the larger is the upper size limit. For EOs associated with early-
   and late-type galaxies, the EO luminosity functions peak at $-$6.40 mag 
   and $-$6.47, respectively, which is about one magnitude fainter than the peak of the GC 
   luminosity function. EOs and GCs form a 
   coherent structure in the $r_{\rm eff}$ vs. $M_{\rm V}$ parameter space, while there is a clear
   gap between EOs and early type dwarf galaxies. However, there is a small potential overlap
   at the high-mass end, where the most extended EOs are close to the parameters of some compact 
   elliptical galaxies. We compare the EO sample with the numerical models of a previous paper and
   conclude that the parameters of the EO sample as a whole can be very well explained by a star cluster 
   origin, where EOs are the results of merged star clusters of cluster complexes (CCs).}
   {}


   \maketitle
%

\section{Introduction}
Globular clusters (GCs) are typically very old stellar objects with masses between 
$10^4$ M$_{\sun}$ and $10^6$ M$_{\sun}$ (corresponding roughly to total luminosities between 
$M_{\rm V} = -5$ to $M_{\rm V} = -10$ mag), having in general compact sizes with half-light radii of a few parsecs. 
This morphology makes them easily observable also in external galaxies with modern telescopes 
\citep[see][and references therein]{brodie06}.

The Milky Way has a rich GC system containing 157 GCs \citep[][2010 edition]{harris}. Most of them 
are compact with sizes of a few parsec. Only 13 GCs (or 8\%) have an effective 
radius larger than 10~pc. 
Most of these extended clusters (ECs) are fainter than $M_{\rm V} = -7$ mag, only NGC2419, 
having a half-light radius of about 20~pc, has a high luminosity of about 
$M_{\rm V} = -9.4$ mag. Two of the 13 ECs, Arp2 and Terzan8, were classified as Milky Way clusters in 
\citet[][2010 edition]{harris}, but the proximity to the Sagittarius dwarf spheriodal galaxy suggests that 
they are related to this Milky Way satellite \citep{salinas}.
Further ECs in the vicinity of the Milky Way have been found in the 
LMC and the Fornax dwarf galaxy \citep{mackey04,vandenbergh04,mvdm}.

Comparable objects have also been detected around other galaxies. 
\citet{huxor04} found three ECs around M31, which have very large radii above 30~pc. 
These clusters were detected by chance as the automatic detection algorithms of the 
MegaCam Survey discarded such extended objects as likely background contaminations. 
Further observations increased the number of ECs in M31 to 13 \citep{huxor08}. 
However, \cite{huxor11} showed that the previous estimates of the effective radii were
considerably too large. The new size estimates are well below 30~pc. 

In addition to ECs located in galactic halos, \cite{larbro00} and \cite{bro02} have discovered a 
population of ECs co-rotating with the disk of the lenticular galaxy NGC\,1023. These so-called faint 
fuzzies have similar structural parameters as halo ECs.

\citet{chandar04} observed a part of the disks of the nearby spiral galaxies M81, M83, 
NGC6946, M101, and M51 using HST and found ECs with effective radii 
larger than 10 pc in four of them. In recent years, hundreds of ECs have been detected in all types 
of galaxies ranging from dwarfs to giant elliptical galaxies.
\begin{table}
\caption{Catalog of the 813 EOs presented in this paper. } 
\label{table_EO-cat} 
\centering 
\begin{tabular}{llcccl} 
\hline\hline 
 & name & r$_{\rm eff}$  & M$_{\rm V}$ & R$_{\rm proj}$ & Ref.\\
 &      &(pc)   &(mag)  &  (kpc)       & \\
\hline 
1 & MilkyWayEO-01 & 21.4 & -9.42 & 89.9	& 1	\\
2 & MilkyWayEO-02 & 10.7 & -6.98 & 16.3	& 1	\\
3 & MilkyWayEO-03 & 13.2 & -6.76 & 17.8	& 1	\\
... &&&&&\\
811 & ESO325-G004EO-13 & 60.8 &	-11.53 & 24.4 &	51	\\
812 & ESO325-G004EO-14 & 67.1 &	-11.46 & 31.6 &	51	\\
813 & ESO325-G004EO-15 & 44.1 &	-11.34 & 54.5 &	51	\\
\hline 
\end{tabular}
\tablefoot{This table is available in its entirety in a machine-readable form at the CDS. 
A portion is shown here for guidance regarding its form and content. The columns 
denote 1. running number, 2. designation of EO in this 
paper, 3. effective radius of the EO, 4. absolute V-band luminosity of the EO, 
5. projected distance to the host galaxy (for the Milky Way the galacto-centric 
distance is used), 6. references for the data: 
(1) \cite{harris}, (2) \cite{salinas}, (3) \cite{vandenbergh04}, (4) \cite{hwang11}, (5) \cite{huxor08},
(6) \cite{huxor11}, (7) \cite{peacock}, (8) \cite{stonkute}, (9) \cite{cockcroft11}, (10) \cite{georgiev}, 
(11) \cite{taylor}, (12) \cite{gomez}, (13) \cite{mclaughlin}, (14) \cite{mouhcine}, (15) \cite{chattopadhyay}, 
(16) \cite{nantais}, (17) \cite{strader}, (18) \cite{sharina05}, (19) \cite{vandenbergh06}, (20) \cite{dacosta}, 
(21) \cite{chandar04}, (22) \cite{hwang08}, (23) \cite{hau}, (24) \cite{lar01b}, (25) \cite{harris09}, 
(26) \cite{larbro00}, (27) \cite{norris11}, (28) \cite{chies11}, (29) \cite{hasegan}, (30) \cite{chilingarianmamon11}, 
(31) \cite{evstigneeva08}, (32) \cite{brodie11}, (33) \cite{evstigneeva07}, (34) \cite{chisa}, (35) \cite{mieske08}, 
(36) \cite{hilker07}, (37) \cite{richtler}, (38) \cite{chilingarian11}, (39) \cite{degraaf}, (40) \cite{goudfrooij}, 
(41) \cite{blom}, (42) \cite{chies06}, (43) \cite{cantiello09}, (44) \cite{darocha10}, (45) \cite{mieske07}, 
(46) \cite{misgeld11}, (47) \cite{penny}, (48) \cite{chiboucas11}, (49) \cite{madrid}, (50) \cite{madrid11}, 
(51) \cite{blakeslee}}. 
\end{table}

\citet{hilker99} and \citet{drinkwater00} discovered in the Fornax Cluster compact objects 
with luminosities above the brightest known GCs and which were not resolved by ground-based 
observations. These objects have masses between a few $10^{6}$~M$_{\sun}$ and $10^{8}$~M$_{\sun}$ 
and effective radii between $r_{\rm eff} = 10$ and 100~pc. 
\cite{phillipps} interpreted these objects as a new type of galaxy and reflected this 
notion in the name ``ultra-compact dwarf galaxy'' (UCD). 
Next to the Fornax Cluster, many UCDs have been found also in other clusters like the 
Virgo Cluster \citep{hasegan,evstigneeva07}, the Centaurus Cluster \citep{mieske07}, 
the Coma Cluster \citep{madrid}, and the Hydra Cluster \citep{misgeld11}. While most 
known UCDs belong to elliptical galaxies in cluster environments, they have also been 
observed in rather isolated objects like the Sombrero galaxy M104 \citep{hau} or 
the group elliptical NGC3923 \citep{norris11}.  

A number of different origins of UCDs were brought up next to the original proposal of UCDs
being just a new type of galaxy \citep{phillipps}. \citet{bekki01,bekki03} suggested 
that UCDs are the remnants of dwarf galaxies which lost their dark matter halo and 
all stars except for their nucleus. 
Next to the interpretation as a galaxy, UCDs were also considered as high-mass versions 
of normal GCs \citep{mieske02}, or as merged massive complexes of star clusters 
\citep{krou98,fellhauer02a,bekki04}. \citet{forbes08} and \citet{mieske08} analyzed the parameters 
of UCDs and concluded that UCDs are more likely bright extended clusters than naked cores of stripped 
dwarf galaxies. The marginally enhanced mass-to-light ratios of UCDs can be explained by slightly 
modified initial stellar mass functions \citep{miekrou08,dabringhausen09}.                                                                                                      

High-resolution HST imaging of gas-rich galaxies experiencing major interactions 
has resolved very intense star formation bursts. \citet{bastian06} observed star cluster 
complexes (CCs), i.e. clusters of young massive star clusters, in the Antennae with masses 
of the order $\approx 10^{6}$~M$_{\sun}$ and diameters of the order 100 to 200 pc. 
\citet{whitmore05} found that the cluster to cluster velocity dispersion in some CCs in the 
Antennae, aka the knots, is small enough to keep them gravitationally bound. \citet{pellerin} 
detected young massive CCs with masses between $10^6$ M$_{\sun}$ and $10^{7.5}$ M$_{\sun}$ and 
diameters between 600 pc and 1200 pc in the collisional ring galaxy NGC\,922. One of the most 
extended CCs has been observed by \cite{tran} in the tail of the ``Tadpole galaxy'' 
UGC\,10214. This CC, which has a mass of the order $10^6$ M$_{\sun}$, has an effective 
radius of 160 pc and a diameter of about 1500 pc.
In a previous paper, we proposed a common origin of ECs and UCDs based on the merged
star cluster scenario \citep{bruens11}. We performed a parametric study that systematically 
scanned a suitable parameter space of CCs and performed numerical simulations to study their 
further evolution. We concluded that the observed ECs and UCDs can be well explained as evolved star CCs.
\begin{table*}
\caption{Catalog of the 65 galaxies containing EOs. } 
\label{table_EO-cat_gal} 
\centering 
\begin{tabular}{lccccccccc} 
\hline\hline 
Galaxy & Type & M$_{\rm V,gal}$ & D$_{\rm gal}$ & Pix$_{\rm HST}$ & N$_{\rm EO}$ & M$_{\rm V,EO,min}$ & M$_{\rm V,EO,max}$ & N$_{\rm GC}$ & Ref \\
       &      & (mag)           & (kpc)         & (pc)            &              & (mag) 	      & (mag) & & \\
\hline
Milky Way &	LT	&	-20.5	&	0	&		&	11	&	-4.73	&	-9.42	&	142	& 1	\\
SAGdSph &	LT	&	-13.8	&	27	&		&	2	&	-5.41	&	-5.68	&	2	& 2 \\
LMC	&	LT	&	-18.34	&	50	&		&	4	&	-4.37	&	-7.25	&	12	&	3	\\
Fornax	&	ET	&	-13.3	&	138	&		&	1	&	-5.32	&	-5.32	&	0	&	3	\\
NGC6822	&	LT	&	-16	&	470	&	0.1	&	3	&	-6.06	&	-7.7	&	3	&	4	\\
M31	&	LT	&	-21.8	&	780	&	0.2	&	20	&	-4.4	&	-7.68	&	232	&	5;6;7	\\
M33	&	LT	&	-19.4	&	870	&	0.2	&	2	&	-5.9	&	-6.6	&	4	&	8;9	\\
UGCA86	&	LT	&	-13.55	&	2728	&	0.7	&	1	&	-7.58	&	-7.58	&	11	&	10	\\
UGC8638	&	LT	&	-13.8	&	3285	&	0.8	&	1	&	-6.57	&	-6.57	&	2	&	10	\\
NGC247	&	LT	&	-19.4	&	3636	&	0.9	&	2	&	-6.59	&	-7.42	&	0	&	10	\\
NGC5128	&	ET	&	-21.4	&	3676	&	0.9	&	26	&	-5.2	&	-11.17	&	194	& 11;12;13;14;15	\\
M81	&	LT	&	-21.2	&	3692	&	0.9	&	44	&	-4.53	&	-8.59	&	369	&	16	\\
NGC4449	&	LT	&	-18.28	&	3693	&	0.9	&	7	&	-5.35	&	-7.14	&	99	&	17	\\
IKN	&	ET	&	-11.5	&	3680	&	0.9	&	1	&	-6.76	&	-6.76	&	4	&	10	\\
NGC5237	&	LT	&	-15.7	&	3794	&	0.9	&	1	&	-6.85	&	-6.85	&	2	&	10	\\
ESO269-58 &	LT	&	-16.3	&	3825	&	0.9	&	2	&	-6.62	&	-6.86	&	6	&	10	\\
UGC7605	&	LT	&	-13.8	&	4177	&	1	&	1	&	-6.44	&	-6.44	&	0	&	18;19	\\
Scl-dE1	&	ET	&	-11.1	&	4300	&	1	&	1	&	-6.7	&	-6.7	&	0	&	20	\\
KK065	&	LT	&	-13.32	&	4510	&	1.1	&	1	&	-6.75	&	-6.75	&	0	&	18;19	\\
NGC784	&	LT	&	-17.6	&	4560	&	1.1	&	2	&	-6.4	&	-6.62	&	5	&	10	\\
M83	&	LT	&	-21	&	4659	&	1.1	&	1	&	-8.24	&	-8.24	&	20	&	21	\\
NGC4605	&	LT	&	-18.5	&	4730	&	1.1	&	2	&	-6.34	&	-8.26	&	9	&	10	\\
UGC3974	&	LT	&	-15.2	&	4897	&	1.2	&	1	&	-8.72	&	-8.72	&	4	&	10	\\
UGC3755	&	LT	&	-15	&	5166	&	1.3	&	2	&	-6.09	&	-7.54	&	30	&	18;19	\\
KK112	&	LT	&	-12.28	&	5220	&	1.3	&	2	&	-6.21	&	-6.77	&	1	&	18;19	\\
NGC1311	&	LT	&	-16.3	&	5252	&	1.3	&	1	&	-7.33	&	-7.33	&	5	&	10	\\
UGC4115	&	LT	&	-14.12	&	5508	&	1.3	&	1	&	-6	&	-6	&	2	&	18;19	\\
M51	&	LT	&	-21.4	&	8031	&	1.9	&	21	&	-6.91	&	-8.86	&	2203	&	22	\\
M104	&	LT	&	-22.45	&	9000	&	2.2	&	10	&	-6.05	&	-12.3	&	184	&	23;24	\\
NGC891	&	LT	&	-21.2	&	9700	&	2.4	&	6	&	-5.16	&	-7.3	&	37	&	25	\\
KK84	&	ET	&	-14.4	&	10069	&	2.4	&	6	&	-6.64	&	-9.68	&	1	&	18;19	\\
NGC1023	&	ET	&	-21.2	&	11791	&	2.9	&	15	&	-6.02	&	-7.13	&	14	&	26	\\
NGC4546	&	ET	&	-20.9	&	13060	&	3.2	&	1	&	-12.94	&	-12.94	&	0	&	27	\\
NGC4660	&	ET	&	-19.69	&	14875	&	3.6	&	1	&	-8.34	&	-8.34	&	50	&	28	\\
IC3652	&	ET	&	-18.7	&	14960	&	3.6	&	1	&	-11.95	&	-11.95	&	0	&	29	\\
NGC4278	&	ET	&	-20.78	&	15154	&	3.7	&	1	&	-9.93	&	-9.93	&	66	&	28	\\
NGC4486B &	ET	&	-17.64	&	15450	&	3.7	&	1	&	-11.98	&	-11.98	&	0	&	29	\\
M89	&	ET	&	-21.32	&	15574	&	3.8	&	3	&	-10.44	&	-11.6	&	104	&	28;29	\\
M59	&	ET	&	-21.38	&	15821	&	3.8	&	1	&	-13.3	&	-13.3	&	0	&	30	\\
M49	&	ET	&	-22.63	&	16052	&	3.9	&	1	&	-10.47	&	-10.47	&	0	&	29	\\
M86	&	ET	&	-22.21	&	16321	&	4	&	2	&	-9.08	&	-9.66	&	74	&	28	\\
NGC4476	&	ET	&	-18.97	&	16450	&	4	&	1	&	-10.89	&	-10.89	&	0	&	29	\\
M87	&	ET	&	-22.54	&	16675	&	4	&	51	&	-8.1	&	-13.42	&	301	&	28;29;31;32	\\
M85	&	ET	&	-21.98	&	17382	&	4.2	&	4	&	-8.5	&	-11.46	&	55	&	28;29	\\
M84	&	ET	&	-22.12	&	17422	&	4.2	&	1	&	-9.75	&	-9.75	&	92	&	28	\\
NGC1380	&	ET	&	-21.3	&	18221	&	4.4	&	13	&	-5.14	&	-9.45	&	174	&	34	\\
NGC1399	&	ET	&	-21.88	&	18950	&	4.6	&	18	&	-10.02	&	-13.4	&	18	&	31;33;35;36;37;38	\\
NGC1533	&	ET	&	-20.7	&	19400	&	4.7	&	3	&	-7.04	&	-7.52	&	136	&	39\\
NGC3923	&	ET	&	-21.9	&	21280	&	5.2	&	3	&	-11.29	&	-12.43	&	0	&	27	\\
NGC1316	&	ET	&	-22.91	&	21900	&	5.3	&	45	&	-5.65	&	-9.26	&	433	&	40	\\
NGC4365	&	ET	&	-22.13	&	23100	&	5.6	&	217	&	-4.66	&	-11.84	&	2038	&	41	\\
NGC5846	&	ET	&	-22.18	&	26709	&	6.5	&	13	&	-6.89	&	-9.32	&	41	&	42	\\
NGC3370	&	LT	&	-20	&	27376	&	6.6	&	22	&	-5.51	&	-7.61	&	255	&	43	\\
NGC1199	&	ET	&	-21.25	&	33100	&	8	&	2	&	-11.97	&	-12.15	&	8	&	44	\\
NGC4696	&	ET	&	-22.81	&	37582	&	9.1	&	2	&	-11.1	&	-11.52	&	6	&	45	\\
NGC3311	&	ET	&	-22.2	&	47200	&	11.4	&	19	&	-9.29	&	-13.35	&	7	&	46	\\
NGC1275	&	ET	&	-22.72	&	71000	&	17.2	&	84	&	-9.98	&	-13.33	&	0	&	47	\\
IC4041	&	ET	&	-20.74	&	94400	&	22.9	&	4	&	-10.77	&	-11.31	&	0	&	48	\\
NGC4889	&	ET	&	-23.51	&	94400	&	22.9	&	2	&	-11.11	&	-11.34	&	2	&	48	\\
IC3998	&	ET	&	-20.43	&	98950	&	24	&	1	&	-11.79	&	-11.79	&	0	&	48	\\
IC4030	&	ET	&	-19.6	&	98950	&	24	&	2	&	-10.7	&	-11.41	&	0	&	48	\\
NGC4874	&	ET	&	-23.25	&	98950	&	24	&	36	&	-9.85	&	-14.03	&	2	&	48;49	\\
NGC1132	&	ET	&	-22.65	&	99500	&	24.1	&	39	&	-9.24	&	-14.8	&	11	&	50	\\
NGC4908	&	ET	&	-21.82	&	101114	&	24.5	&	3	&	-11.08	&	-12.47	&	0	&	48	\\
ESO325-G004	&	ET	&	-23.2	&	143000	&	34.7	&	15	&	-11.34	&	-13.51	&	0	&	51	\\
\hline 
\end{tabular}
\tablefoot{The columns denote 1. name of the galaxy, 2. type of host galaxy (LT: late-type, ET: early-type),
3. absolute V-band luminosity of the galaxy, 4. distance of the galaxy, 5. size of a HST ACS pixel at the
distance of the galaxy, 6. number of EOs, 7. minimum absolute V-band luminosity of the EO, 
8. maximum absolute V-band luminosity of the EO, 9. number of GCs in the same publications,
10. references as in Table \ref{table_EO-cat}.}
\end{table*}

In recent years, the number of observed ECs and UCDs has rapidly increased for all types of galaxies 
in various environments. As there is no clear distinction between ECs and UCDs, both types of objects 
will be called extended stellar objects -- abbreviated "EOs" -- in this paper. The high number of EOs
known today, allows for the first time for a detailed analysis of the properties of EOs to disclose 
commonalities and distinctions of objects from early- and late-type galaxies.

The paper is structured as follows: in Section \ref{observations} we compile a catalog of ECs and UCDs 
on the basis of the available publications containing structural parameters of ECs and UCDs.
In Section \ref{results} we present the results of the catalog that are discussed in Section \ref{discussion}.
Section \ref{summary} provides a summary and conclusions.

\section{Observational basis}\label{observations}
As already indicated in the previous section, GC-like objects with effective radii above 
10 pc, which cover a large luminosity range, have been found in various environments 
from dwarf to giant elliptical galaxies. To allow 
for an analysis of their parameters, we compiled a catalog of effective radii, absolute V-band 
luminosities, and projected distances of EOs to their host galaxies as well 
as the absolute V-band luminosities of these galaxies and their distance to the Milky Way. 
We distinguish between EOs found in late-type galaxies, i.e. spiral and 
irregular galaxies, and early-type galaxies, i.e. elliptical, lenticular, and dwarf spheroidal galaxies.
In Table \ref{table_EO-cat}, available at the CDS, we present the catalog of the 813 
EOs used in this study. Table \ref{table_EO-cat_gal} provides an overview on the galaxies, 
where EOs were detected, the number of EOs per galaxy and the luminosity range of the detected EOs.

\subsection{EOs in Late-Type Galaxies}
According to the 2010 edition of the GC catalog of \cite{harris} and considering that Arp2 and Terzan8 are
associated with the Sagittarius dwarf spheriodal galaxy \citep{salinas}, the Milky Way has 11 EOs. 
The other two Local Group spiral galaxies M31 and M33 have 20 \citep{huxor08,peacock,huxor11} and 
2 \citep{stonkute,cockcroft11} EOs, respectively. EOs were also found in the LMC \citep{vandenbergh04}, 
and the dwarf irregular galaxy NGC6822 \citep{hwang11}.

Outside the Local Group, EOs were detected in the spiral galaxies M81 \citep{chandar04,nantais}, 
M83 \citep{chandar04}, M51 \citep{chandar04,hwang08}, NGC891 \citep{harris09}, 
NGC3370 \citep{cantiello09}, and in the Sombrero Galaxy M104 \citep{lar01b,hau}.
In addition, EOs were found in the dwarf irregular and the Magellanic type dwarf galaxies UGCA86, UGC8638, 
NGC247, NGC5237, ESO269-58, NGC784, NGC4605, UGC3974, and NGC1311 \citep{georgiev}, UGC7605, KK065, UGC3755, 
KK112, and UGC4115 \citep{sharina05,vandenbergh06}, and NGC4449 \citep{strader}. 

In total, the EO catalog contains 171 EOs associated with late-type galaxies.

\begin{figure}[t]
\centering
\includegraphics[width=8.8cm]{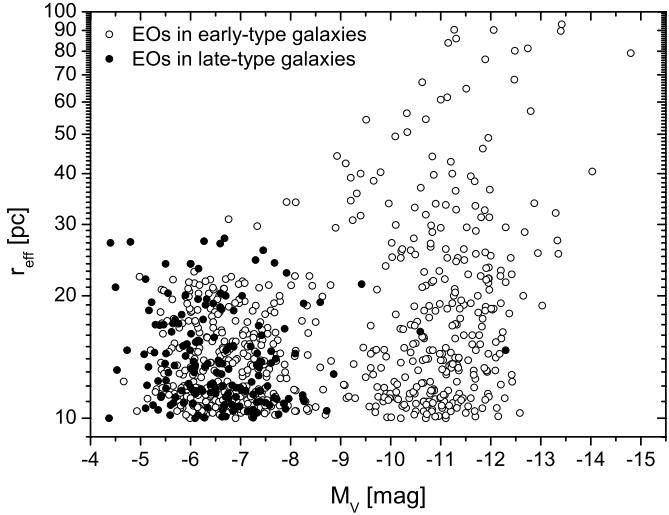}
\caption{Effective radii of EOs are plotted against their absolute
V-band luminosities. EOs associated with early-type galaxies are plotted as open circles, EOs 
associated with late-type galaxies are plotted as black circles.}
\label{fig_reffmv}
\end{figure}

\subsection{EOs in Early-Type Galaxies}
EOs were detected in a large number of elliptical galaxies: 
NGC5128 \citep{gomez,mclaughlin,chattopadhyay,taylor,mouhcine}, NGC4660 \citep{chies11}, IC3652 \citep{hasegan}, 
NGC4278 \citep{chies11}, NGC4486B \citep{hasegan}, M89 \citep{hasegan,chies11}, M59 \citep{chilingarianmamon11}, 
M49 \citep{hasegan}, M86 \citep{chies11}, the central galaxy of the Virgo Cluster M87 
\citep{hasegan,evstigneeva08,brodie11,chies11}, M84 \citep{chies11}, 
the central galaxy of the Fornax Cluster NGC1399
\citep{richtler,evstigneeva07,hilker07,evstigneeva08,mieske08,chilingarian11}, NGC3923 \citep{norris11}, NGC4476 \citep{hasegan}, 
NGC5846 \citep{chies06}, NGC4696 \citep{mieske07}, NGC3311 \citep{misgeld11}, IC4041, NGC4889, IC3998, IC4030, 
IC4041, and NGC4908 \citep{chiboucas11}, NGC4874 \citep{madrid,chiboucas11}, NGC1132 \citep{madrid11}, NGC1275
\citep{penny}, NGC4365 \citep{blom}, NGC1316 \citep{goudfrooij}, NGC1199 \citep{darocha10}, 
and ESA325-G004 \citep{blakeslee}.

In five lenticular galaxies, EOs were detected: in NGC1023 \citep{larbro00,bro02} 
and NGC1380 \citep{chisa}, 15 and 13 EOs were found, respectively. One EO was discovered in NGC4546 
\citep{norris11}, three EOs in NGC1533 \citep{degraaf}, and four EOs were found in M85 \citep{hasegan,chies11}. 

The EO of the dwarf elliptical Scl-dE1 \citep{dacosta}, the two EOs of the Sagittarius dSph galaxy \citep{salinas}, 
the EO of the dSph galaxy Fornax \citep{vandenbergh04}, 
the 6 EOs of the dSph galaxy KK84 \citep{sharina05,vandenbergh06}, and the EO of the dSph galaxy IKN \citep{georgiev} 
are contributing to the list of EOs associated with early-type galaxies. 

In total, 642 EOs were found in early-type galaxies, 595 thereof are associated with elliptical galaxies.
The EO sample of the elliptical galaxies is dominated by the two galaxies NGC4365 and NGC1275, which
have 217 and 84 EO candidates, respectively.
\begin{figure*}
\centering
\includegraphics[width=18.2cm]{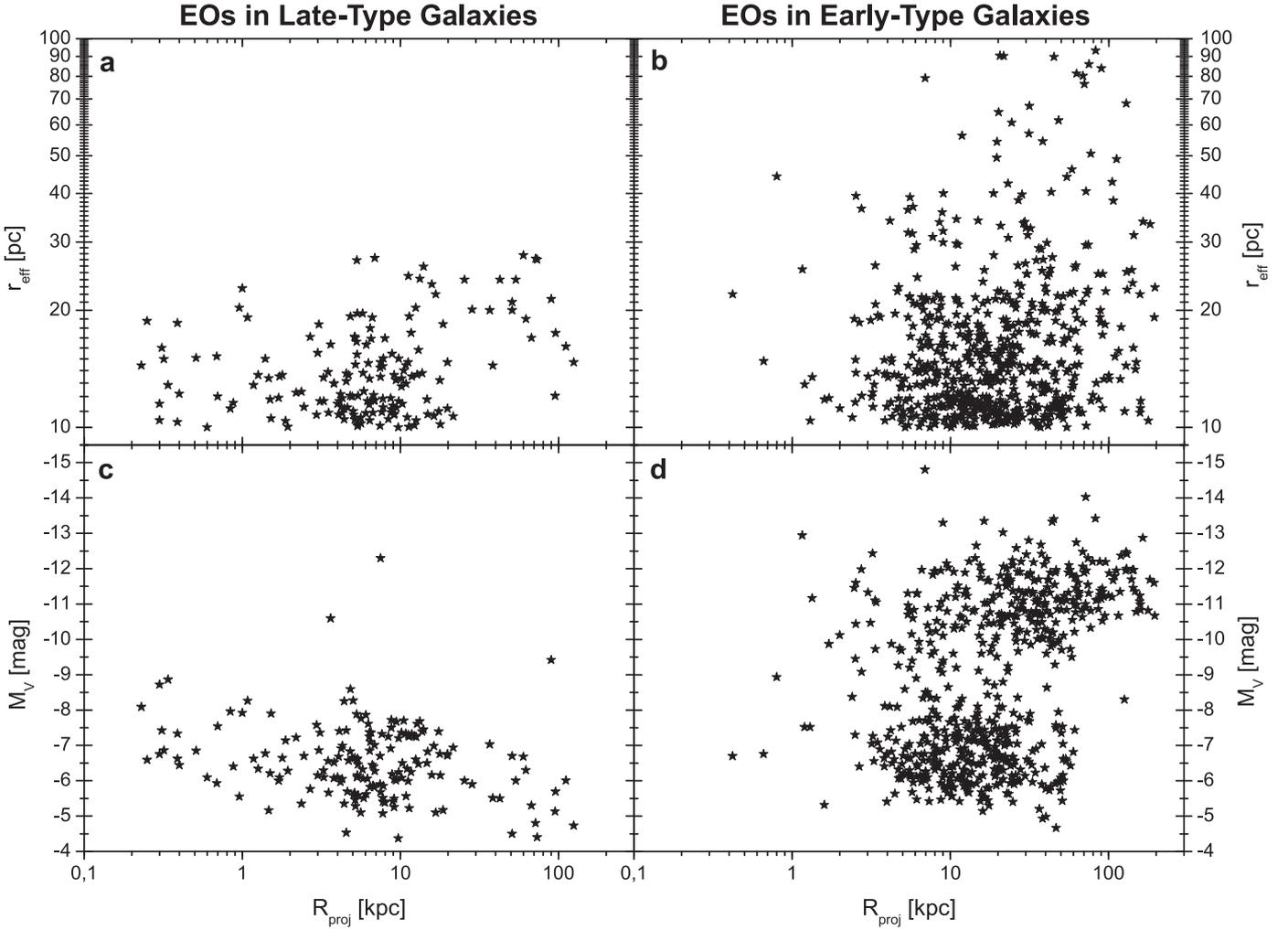}
\caption{(a) and (b) Effective radii of EOs are plotted against the projected 
distances to their host galaxies. 
(c) and (d) Absolute V-magnitudes of EOs are plotted against the projected distance.}
\label{fig_rproj}
\end{figure*}

\section{Results}\label{results}

\subsection{Correlations between the Parameters of Extended Objects}
Figure~\ref{fig_reffmv} shows the effective radii $r_{\rm eff}$ of the 813 EOs of our catalog
as a function of their total V-band luminosities $M_{\rm V}$.
The vast majority of EOs associated with late-type galaxies have magnitudes between
$M_{\rm V}=-4$ to $-9$ mag. Only three EOs, the Milky Way cluster NGC2419 and two EOs
associated with the Sombrero Galaxy M104, have total V-band luminosities brighter than $M_{\rm V}=-9$ mag.
EOs associated with early-type galaxies cover a significantly larger range of V-band luminosities.
The majority of objects are found in the magnitude range of about $M_{\rm V}=-5$ to $-13$ mag. 
At $M_{\rm V}=-8.5$ mag the number of objects is much smaller than for lower and higher luminosities. 

At each magnitude EOs are found with effective radii between 10 pc and an upper size limit, which 
shows a clear trend: the more luminous the object the larger is the upper size limit.

The dependency of the structural parameters effective radius, $r_{\rm eff}$, and luminosity, $M_{\rm V}$, 
of EOs on the projected distance from their host galaxy, $R_{\rm proj}$, are displayed in Figure \ref{fig_rproj}. 
For late-type galaxies there is a slight trend of increasing effective radii and decreasing total luminosities
with increasing projected distance (Fig. \ref{fig_rproj}a and c). 
On the other hand, very faint and very extended objects are extremely hard to detect at very
low projected distances due to the high surface brightness of the underlying host galaxy.
Consequently, the slight trends might not be significant.
In early-type galaxies the most extended objects are found predominantly at large 
projected distances (Fig. \ref{fig_rproj}b). The most extended EO, VUCD7, 
with an effective radius of 93.2 pc, was discovered in the central elliptical galaxy M87 of the Virgo 
Cluster at a projected distance of 82.6 kpc. 
In early-type galaxies, the low-luminosity objects are comparable to those 
in late-type galaxies and there are high-luminosity objects that are not present in late-type galaxies. 
The void at projected
distances larger than about 70 kpc and luminosities fainter than -10.5 mag is a result of
the limited coverage and sensitivity of most surveys.

\begin{figure}[t]
\centering
\includegraphics[width=8.9cm]{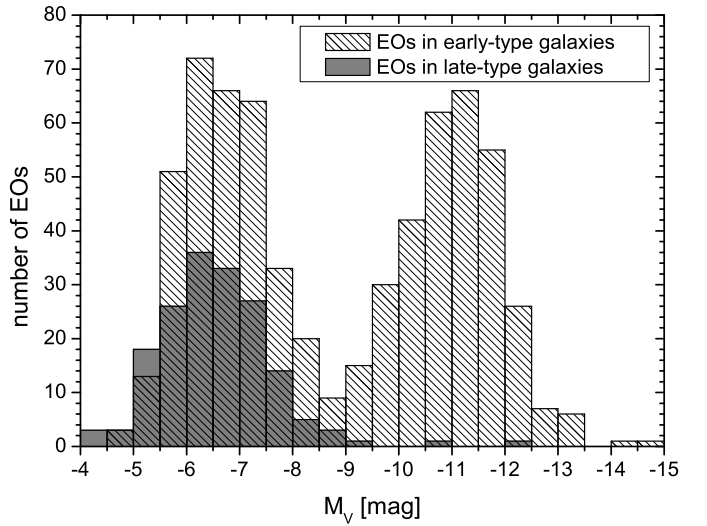}
\caption{Histogram of the number of EOs in early-type and late-type galaxies at different 
total V-band luminosities.}
\label{fig_histmvall}
\end{figure}

Figure \ref{fig_histmvall} shows two histograms of the number of EOs with different total 
V-magnitudes. The EOs in late-type galaxies have a luminosity distribution which peaks at around 
$M_{\rm V}= -6.5$ mag, while the EOs in early-type galaxies show a bimodal distribution which 
peaks at $M_{\rm V}= -6.5$ mag and $-$11.0 mag and has a clear minimum between $-$8.5 and $-$9 mag. 
In Section \ref{sectlumiearly}, it will be shown that the bimodal luminosity distribution 
is mostly a selection effect due to EO samples covering solely the high luminosity regime. 

\begin{figure}[t]
\centering
\includegraphics[width=8.9cm]{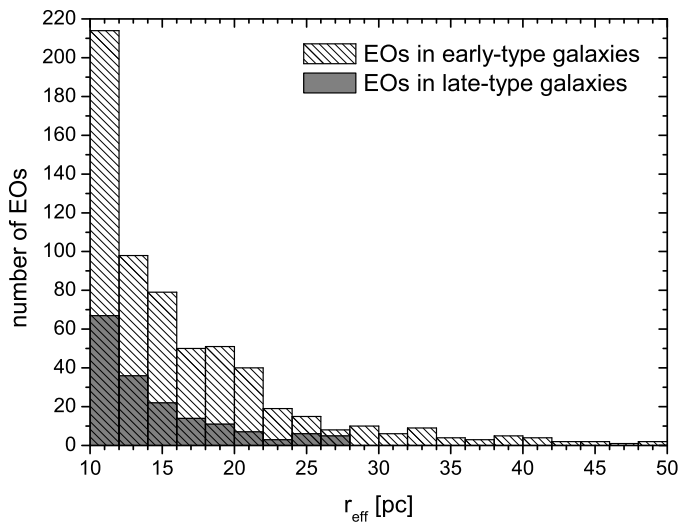}
\caption{Histogram of the number of EOs in early-type and late-type galaxies at different 
effective radii.}
\label{fig_histreff}
\end{figure}

\begin{figure}[t]
\centering
\includegraphics[width=8.9cm]{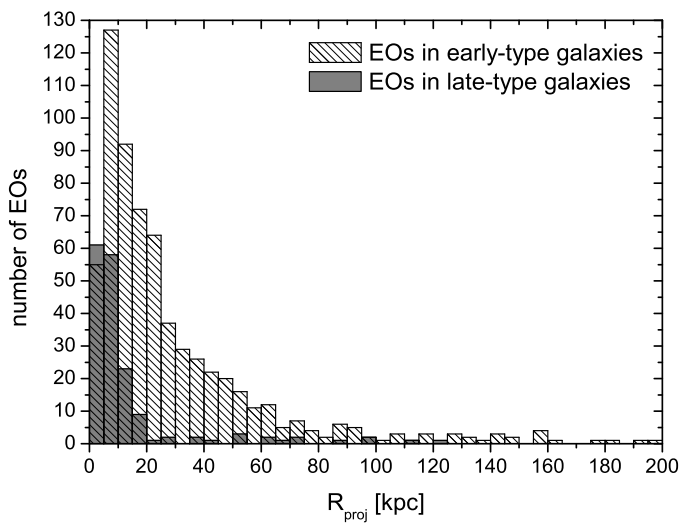}
\caption{Histogram of the number of EOs in early-type and late-type galaxies at different 
projected radii.}
\label{fig_histrpro}
\end{figure}

Figure \ref{fig_histreff} shows two histograms of the number of EOs with different effective radii.
For EOs in both early- and late-type galaxies, the majority of objects 
has small effective radii. The mean size for EOs in early-type galaxies is 18.1 pc and its median 
value lies at 14.2 pc. 
For EOs in early-type galaxies that are fainter than $M_{\rm V}= -10$ mag, the mean and the median
size are 15.2 pc and 13.5 pc, respectively.
The sizes of EOs in late-type galaxies are slightly smaller: the mean size is 14.4 pc 
and its median value is 13.2 pc. 
For EOs in late-type galaxies, 39.2\% have effective radii between 10 and 12 pc. For early-type
galaxies, 38.8\% and 25.0\% are in the interval between 10 and 12 pc for EOs fainter and brighter
than $M_{\rm V}= -10$ mag, respectively.

A histogram of the projected distance of EOs to their host galaxies is presented in Figure \ref{fig_histrpro}. 
EOs in late-type galaxies were predominantly found at small projected distances below 20 kpc. It should
be noted, however, that the coverage of the halo beyond projected distances of 20 kpc is extremely poor
for most late-type galaxies.
EOs in early-type galaxies were found also at considerably larger distances. The mean and the median
projected distances are 12.9 kpc and 6.4 kpc for EOs in late-type galaxies and 28.7 kpc and 17.9 kpc 
for EOs in early-type galaxies, respectively. Only few EOs have been discovered beyond 100 kpc. All 
EOs with projected distances larger than 130 kpc are associated with the giant elliptical galaxies in 
the center of the Virgo Cluster, the Fornax Cluster, the Perseus Cluster, and the Coma Cluster.
This result does not necessarily imply that only the central galaxies of clusters have EOs at
very large distances, as rather isolated galaxies scarcely have HST observations at such 
large projected distances (see Sect.~\ref{incomplete_coverage}).

\subsection{Correlation of EO parameters with those of their host galaxies.}

\begin{figure*}
\centering
\includegraphics[width=18.2cm]{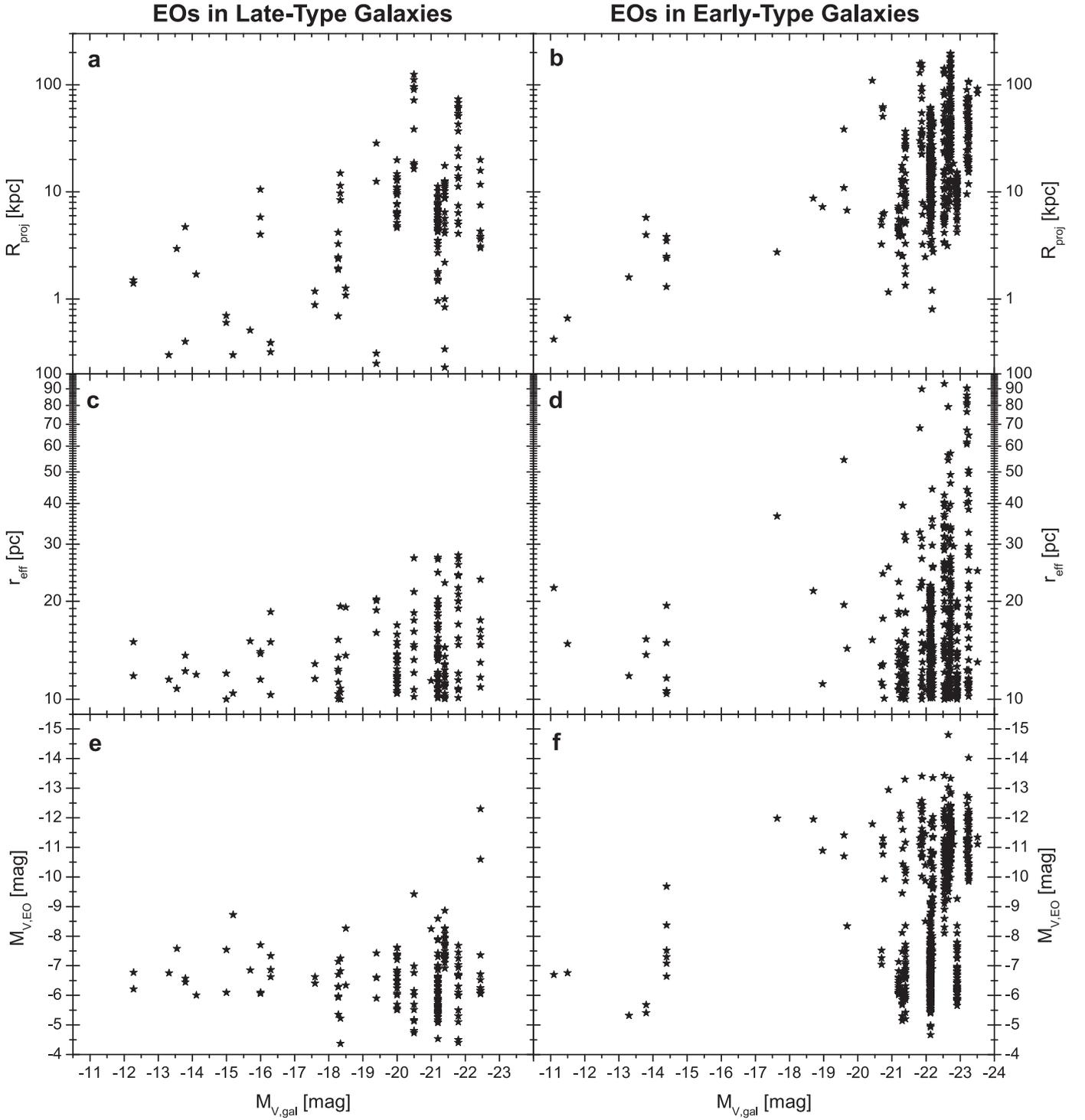}
\caption{(a) and (b) Projected distances of EOs are plotted against the total 
V-band luminosity of their host galaxies.
(c) and (d) Effective radii of EOs are plotted against the total V-band luminosity of the host galaxies.
(e) and (f) Absolute V-magnitudes of EOs are plotted against the total V-band luminosity of the host galaxies.}
\label{fig_mvgal}
\end{figure*}

Figure \ref{fig_mvgal} shows the dependence of the parameters projected distance of EOs to their host galaxies, 
$R_{\rm proj}$, effective radius, $r_{\rm eff}$, and absolute V-band magnitude of the EOs, $M_{\rm V,EO}$, on 
the total V-band luminosity of the host galaxies, $M_{\rm V,gal}$. 

Figures \ref{fig_mvgal}a and b illustrate the dependence of the projected distances of the EOs on the 
total V-band luminosity of the host galaxies. For EOs in late-type galaxies only three 
galaxies (the Local Group galaxies Milky Way, M31, and M33) host EOs at projected distances beyond 20 kpc.
In contrast, for early-type galaxies with luminosities between $-$21 and $-$24 mag EOs are found far out 
from the galactic center even beyond 100 kpc. Only very few objects are found at projected distances 
less than a few kiloparsecs. This lack of EOs at small projected distances is at least partly
due to the fact that EOs have a very low contrast on the bright background light from the host galaxy.

Figures \ref{fig_mvgal}c and d demonstrate that the upper limit of the effective radii of EOs in 
late-type and early-type galaxies tends to increase with the total V-band luminosity of the host galaxy. 
For all galaxies, most EOs have effective radii well below 20 pc.

Figure \ref{fig_mvgal}e shows that the total luminosity of EOs in late-type galaxies does not depend on 
the total luminosity of their host galaxy. For EOs in early-type galaxies (Fig. \ref{fig_mvgal}f) 
there is a trend: the more massive the parent galaxy the higher is the upper limit of the luminosities of its EOs.
This trend is partly a size-of-sample effect, as in large EO samples also extreme luminosities 
can be realized. In Section \ref{sectlumiearly} we will demonstrate, however, that the number of high-luminosity
objects is larger than expected from the EO luminosity function.

\begin{figure}[t]
\centering
\includegraphics[width=8.8cm]{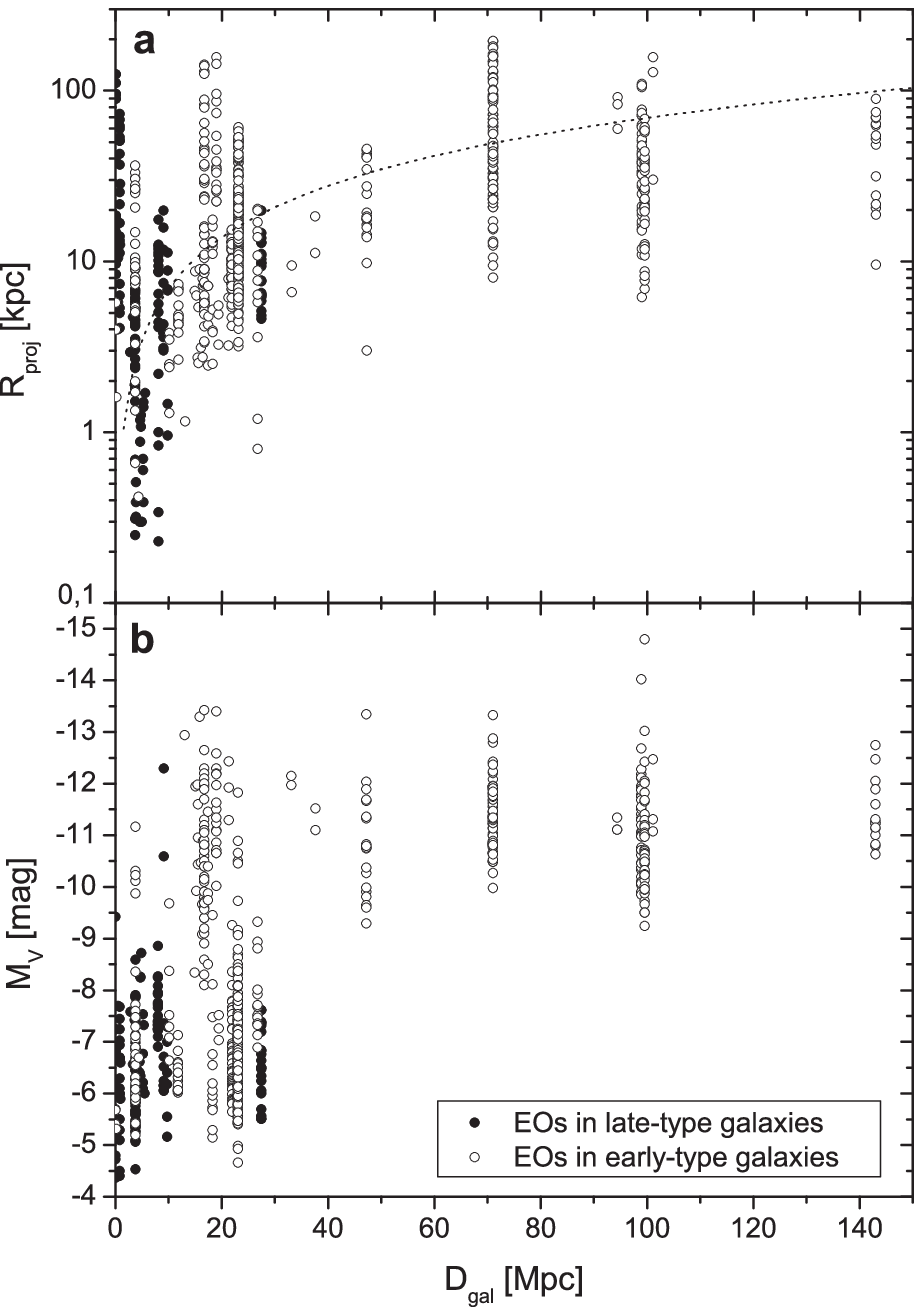}
\caption{(a) Projected distances of EOs from their host galaxies are plotted against the distance 
of the host galaxies. The dotted curve indicates the largest possible projected radius within a single HST ACS 
image centered on a galaxy at a given distance. (b) Absolute V-magnitudes of EOs are plotted against the 
distance of the host galaxies.}
\label{fig_dgal}
\end{figure}

Figure \ref{fig_dgal}a plots the projected distances of the EOs versus the distance of their 
host galaxies from the Galaxy. Early-type galaxies have been studied nearby as well as at large distances 
up to 143 Mpc, while late-type galaxies were mainly observed in the local universe up to distances of 10 Mpc. 
Only one spiral (NGC\,3370) was searched for EOs at a larger distance of about 27 Mpc.

Figure \ref{fig_dgal}b displays the absolute V-magnitude of the EOs versus the distance of 
the host galaxy from the Galaxy. The maximum absolute V-magnitude of EOs does not change 
with the distance of the host galaxy. While EOs brighter than $M_{\rm V}$ = $-$10 mag were detected at all
distances, faint EOs have only been observed in galaxies closer than 30 Mpc.

\section{Discussion}\label{discussion}

\subsection{Distribution in the $r_{\rm eff}$ vs. $M_{\rm V}$ space.}\label{sect_confirmed}

Figure~\ref{fig_reffmv} shows the effective radii $r_{\rm eff}$ of the 813 EOs as a function of 
their total V-band luminosities $M_{\rm V}$. At all luminosities, EOs cover a range between 10 pc and an 
upper limit, which increases with increasing luminosity from about 25 pc at M$_{\rm V}$ = $-$5 mag to about 100 pc 
at M$_{\rm V}$ = $-$13 mag.

\begin{figure}[t]
\centering
\includegraphics[width=8.8cm]{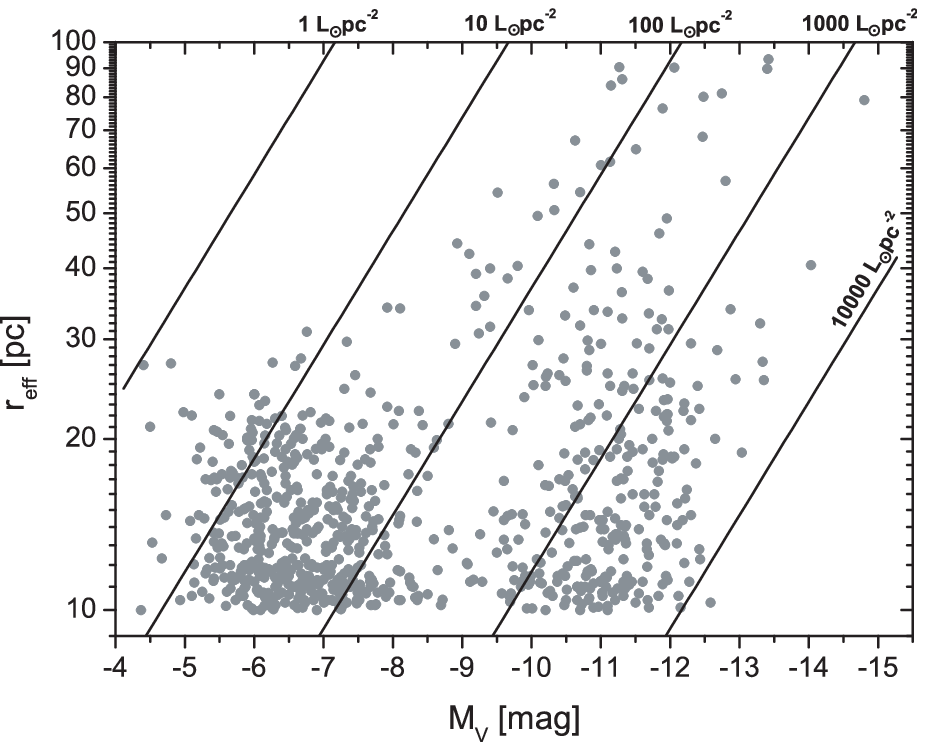}
\caption{Effective radii of EOs are plotted against their absolute V-magnitudes (grey circles). 
The black lines indicate trends of equal mean surface brightnesses of 1, 10, 100, 1000, and 
10000 L$_{\sun}$ pc$^{-2}$.}
\label{fig_demosurfbright}
\end{figure}

This trend of the increasing upper limit of effective radii with total luminosities 
defines the range where EOs have been found so far. It should not, however, be regarded as 
a firm upper limit, as very extended objects with a low total luminosity are extremely
hard to detect. While the detection limit of EOs depends on numerous parameters like the 
magnitude and the structure of the fore- and background emission, we will focus on the 
characteristics of the EOs themselves. 

The effective radius is defined as the radius where half of the total luminosity of
an object is included. A large effective radius means that the luminosity is spread over
a large area leading to low surface brightness. In order to provide a rough estimate of
the mean surface brightness, we divide the luminosity within the effective radius by the
area of a circle with the size of the effective radius,

\begin{eqnarray} 
\overline{\Sigma_{\mathrm{V,EO}}} = \frac{0.5~L_{\mathrm{V,EO}}}{A(r_{\mathrm{eff}})} = 
\frac{10^{-0.4~(M_\mathrm{V}-M_{\rm V,\sun})}}{2~\pi~r_{\mathrm{eff}}^2}
\label{meansurfbrightness}
\end{eqnarray}

where $\overline{\Sigma_{\mathrm{V,EO}}}$ is the mean surface brightness, $L_{\mathrm{V,EO}}$
is the total luminosity of an EO, $M_{\rm V,\sun} = 4.83$ mag is the absolute solar V-band luminosity, 
and $A(r_{\mathrm{eff}}) = \pi~r_{\mathrm{eff}}^2$ is the area within
$r_{\mathrm{eff}}$. Figure \ref{fig_demosurfbright} shows $r_{\rm eff}$ vs. $M_{\rm V}$ of EOs as in 
Fig.~\ref{fig_reffmv} and in addition lines with constant mean surface brightness as defined in
Eq. \ref{meansurfbrightness}. An EO with $r_{\rm eff}$ = 10 pc and $M_{\rm V}$ = $-$5.0 mag has the
same mean surface brightness as an EO with $r_{\rm eff}$ = 50 pc and $M_{\rm V}$ = $-$8.5 mag.
On the other hand, Fig.~\ref{fig_demosurfbright} demonstrates that the trend of the increasing 
upper limits of effective radii with increasing total luminosity is not aligned with the lines 
of constant mean surface brightness. Consequently, the trend cannot be explained as a simple 
limit of detectability.

\begin{figure}[t]
\centering
\includegraphics[width=8.8cm]{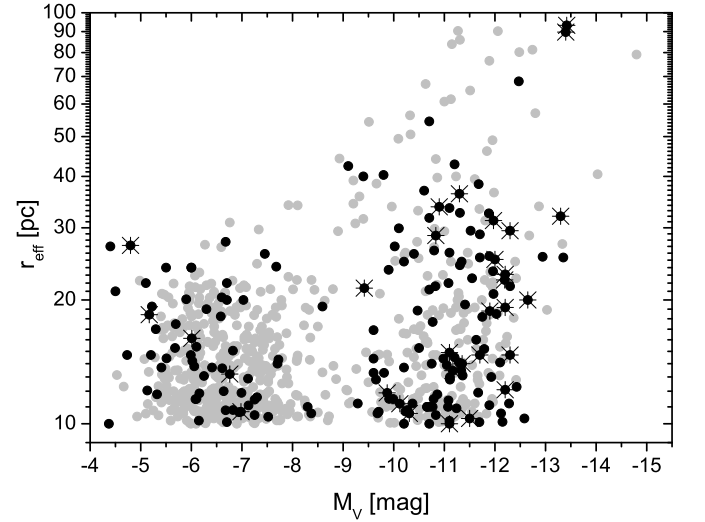}
\caption{Effective radii of EOs are plotted against their absolute V-magnitudes. Black circles indicate
the 175 confirmed EOs, while grey circles are the 638 EO candidates. 
The 31 EOs which have a measured dynamical mass are marked by an additional star.}
\label{fig_confirmedEO}
\end{figure}

The EO catalog and the results presented in Section \ref{results} contain both confirmed EOs and
EO candidates. EOs are identified initially in HST images on the basis of an almost round shape and 
a color consistent with being a GC. The criteria exclude a fair fraction of background galaxies. 
However, the presence of round background galaxies with the same color as GCs cannot be excluded. 
Only in the very neighborhood of the Local Group, EOs can be resolved into stars discriminating them
from background objects. Consequently, all more distant candidates need to be confirmed by follow-on 
spectroscopy measuring radial velocities of the objects. As such a procedure is extremely time consuming, 
only a fraction of the candidate GCs and EOs are observed spectroscopically. 

The large extension of EOs leads to significantly lower central surface brightnesses.
In terms of central surface brightness, an EO with $r_{\rm eff}$ = 10 pc is roughly a factor of 25 or 
3.5 magnitudes fainter than a GC with the same total luminosity, but with $r_{\rm eff}$ = 2 pc. 
An EO with $r_{\rm eff}$ = 30 pc is already a factor of about 225 or 5.9 magnitudes fainter.
Consequently, GCs are typically preferred over EOs during the selection of targets for confirmation, 
as they are considerably easier to confirm.

From the 813 EOs in the catalog, only 175 EOs were so far spectroscopically confirmed. 
Figure \ref{fig_confirmedEO} shows effective radii of confirmed (black) and candidate EOs (grey) as a 
function of their total V-magnitudes. The confirmed EOs cover basically the same M$_{\rm V}$ and $r_{\rm eff}$ 
parameter space as the entire catalog and show the same trend of increasing upper size limits with 
increasing mass. Consequently, the overall distribution of EOs in the $r_{\rm eff}$ vs. $M_{\rm V}$ space
and the trend of increasing upper size-limits with increasing luminosity are not significantly
influenced by contaminating background objects.

\begin{figure*}[t]
\centering
\includegraphics[width=17.9cm]{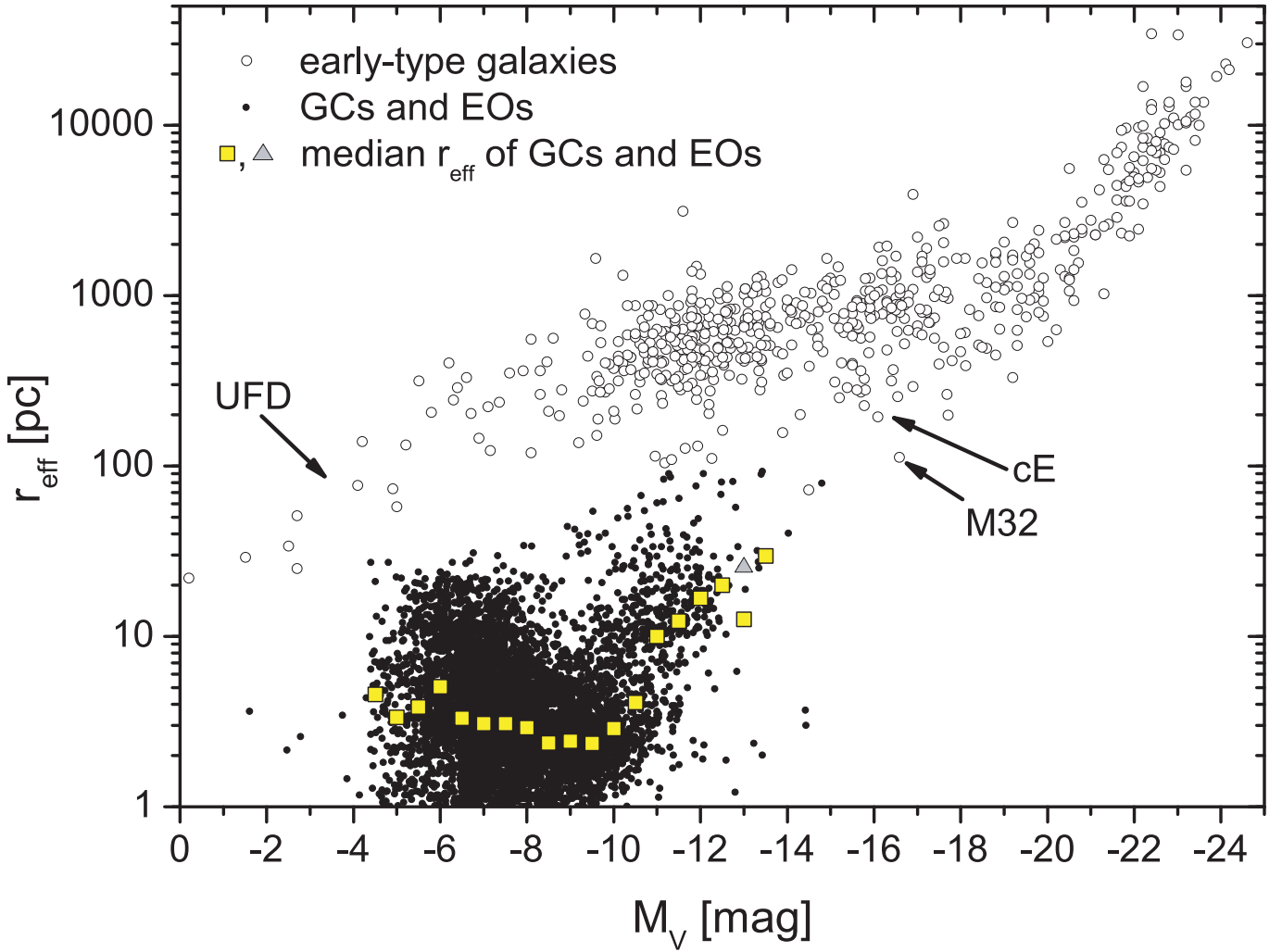}
\caption{Effective radii of GCs and EOs are plotted against their absolute V-magnitudes (black circles).
The median effective radius per luminosity bin is given as squares. 
The six exceptionally bright GCs with luminosities between M$_{\rm V}$ = -12.5 and -14.5 mag and effective radii 
below 4 pc are unconfirmed candidate clusters associated with NGC\,4365 \citep{blom}. The grey 
triangle shows the median effective radius, when these exceptionally bright compact 
GCs were removed in the bin at M$_{\rm V}$ = $-$13.0 mag. In addition, 
early-type galaxies are shown as open circles. Highlighted are the two galaxy types compact ellipticals (cE) 
with their prototype example M32, and the ultra faint dwarf galaxies (UFD) recently found in the Milky Way.}
\label{fig_GCsandEOs}
\end{figure*}

Figure \ref{fig_GCsandEOs} shows next to the effective radii and total luminosities of EOs also
the corresponding parameters of GCs with effective radii smaller than 10 pc. The parameters of the GCs 
were taken from the same papers used for the EOs (see Sect.~\ref{observations}). The diagram demonstrates 
that EOs and GCs form a coherent structure in the M$_{\rm V}$ vs. $r_{\rm eff}$ parameter space. 

\begin{figure}[t]
\centering
\includegraphics[width=8.8cm]{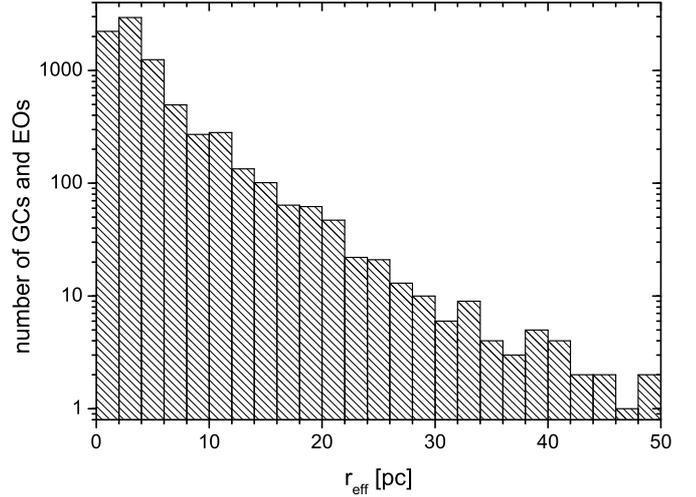}
\caption{Histogram of the number of star clusters at different effective radii for all GCs and EOs
shown in Fig. \ref{fig_GCsandEOs}. The slight increase of numbers at 10 pc is due to the fact
that all publications of this paper were selected to contain EOs, but not all of them also include GCs.}
\label{fig_histreffGCsandEOs}
\end{figure}

Figure \ref{fig_histreffGCsandEOs} is a histogram of the effective radii of all star clusters presented in 
Fig. \ref{fig_GCsandEOs}. The largest number of objects is contained in the bin covering effective radii between
2 and 4 pc. For larger effective radii, the number of objects decreases approximately exponentially. 
Star clusters with effective radii below 6 pc include 80 percent of all objects, while the EOs represent
about 10 percent of the objects. 
 
In Fig. \ref{fig_GCsandEOs}, the median effective radii per luminosity bin of the combined GC-EO-sample are 
given as squares. 
For luminosities fainter than about M$_{\rm V}$ = $-$10.5 mag, compact clusters with effective radii of a few parsec 
dominate. With increasing luminosity, EOs start to dominate over GCs leading to an overall trend of 
increasing effective radii with increasing total luminosity. The median effective radius increases
from 10 pc at M$_{\rm V}$ = $-$11.0 mag to 30 pc at M$_{\rm V}$ = $-$13.5 mag. 
At the high-luminosity end, the number of objects is quite low. The low median effective radius at
M$_{\rm V}$ = $-$13.0 mag is due to three very compact candidate GCs in this bin. The removal of these unconfirmed 
GCs results in a median effective radius that fits the overall trend (grey triangle). 
While the data show a clear trend of increasing effective radii with increasing luminosity, a tight 
size-luminosity relation as seen in older publications \citep[e.g.][]{dabringhausen08, evstigneeva08} is 
no longer existing on the basis of the larger data-set presented in this paper.
This result is consistent with the conclusions of \cite{brodie11}, which were based on a considerably
smaller data-set.

Figure \ref{fig_GCsandEOs} shows next to the GCs and EOs (black circles) also the
effective radii and absolute luminosities of early-type galaxies (open circles). We compiled the parameters 
of the elliptical, dwarf and compact elliptical, and dwarf spheroidal galaxies from \cite{brasseur}, 
\cite{slater}, \cite{bell}, \cite{McConnachie}, \cite{cappellari}, \cite{sharina08}, \cite{mieske05}, 
\cite{huxor11b}, \cite{belokurov}, \cite{misgeldhilker}, \cite{dacosta}, \cite{martin08}, \cite{price09}, 
\cite{geha}, \cite{smithcast08}, \cite{smithcast11}, \cite{blakeslee}, and \cite{McConnachie12}.
The galaxies span a luminosity range of about 24 magnitudes and a size range of three orders of magnitude.

While the star clusters (GCs and EOs) and the early-type galaxies form both a coherent structure in 
the M$_{\rm V}$ vs. $r_{\rm eff}$ parameter space, there is a clear gap between star clusters and galaxies
at least in the luminosity interval between $M_{\rm V}$ = $-$6 mag and $-$11 mag. 
This gap was first discussed by \cite{gilmore}.
At $M_{\rm V}$ = $-$6 mag, EOs have effective radii up to about 30 pc, while the dwarf spheroidal galaxies 
at this luminosity have effective radii between about 100 and 400 pc. In the high-luminosity region at 
about $M_{\rm V}$ = $-$12 mag EOs have effective radii up to about 90 pc, while the dwarf galaxies at 
this luminosity have effective radii between 200 and 1300 pc. 
Between $M_{\rm V}$ = $-$11 and $-$12 mag there are six candidate compact ellipticals from \cite{blakeslee}.
They have the same parameters as EOs, but they are slightly larger than 100 pc, which lead to the 
classification as cEs. As there is no clear distinction between EOs and cEs, these six cEs might as well be
very extended EOs.

The ultra faint dwarf galaxies (UFDs), which were recently found around the Milky Way 
\citep[e.g.][]{martin08,belokurov}, cover the luminosity range between $M_{\rm V}$ = 0 and $-$5 mag.
The detection of large objects with effective radii greater than 20 pc and luminosities fainter
than $M_{\rm V}$ = $-$5 mag is extremely challenging even within the Local Group. While there is no overlap
between UFDs and EOs for the Milky Way, a potential overlap of EOs and UFDs for other galaxies cannot be 
excluded, as neither UFDs nor very faint EOs were within the detection limits of existing surveys.

On the high-luminosity end of the EO distribution, some compact elliptical galaxies have parameters
comparable to the most extended EOs. Figure \ref{fig_GCsandEOs} shows two EOs brighter than $M_{\rm V}$ = $-$14 mag, 
with effective radii of about 40 and 80 pc \citep{madrid,madrid11}.
However, both are so far unconfirmed candidates that might 
be background galaxies. The three most luminous, confirmed objects having $M_{\rm V} \approx$ $-$13.4 mag 
are VUCD7, UCD3, and HUCD1, which are associated with the central elliptical galaxies of the Virgo Cluster, 
M87, of the Fornax Cluster, NGC\,1399, and the Hydra Cluster, NGC\,3311, respectively. In addition, 
the confirmed object M59cO, which is associated with the giant elliptical galaxy M59, has a luminosity of about 
$M_{\rm V} \approx$ $-$13.3 mag. VUCD7 and UCD3 have effective radii of the order of 90 pc, while HUCD1 
and M59cO have effective radii of 25 and 32 pc, respectively.

The Coma Cluster compact elliptical galaxy CcGV19b \citep{price09}, which has a luminosity of 
$M_{\rm V} \approx$ $-$14.5 mag and an effective radius of $r_{\rm eff}$ = 72 pc, is located at a 
projected distance of 68 kpc to NGC\,4874. These parameters could also lead to a classification
of CcGV19b as an EO. The observed mass-to-light ratio of about 13 \citep{price09}, which is about
a factor of three larger than that of VUCD7 and UCD3 \citep{evstigneeva07,mieske08}, suggests
rather a galactic origin. However, the vast majority of EOs, about 96\%, do not have observed velocity 
dispersions (see Fig. \ref{fig_confirmedEO}), which are needed to estimate a dynamical mass. Consequently, 
some overlap between EOs and compact elliptical galaxies cannot be excluded. 

\begin{figure}[t]
\centering
\includegraphics[width=8.8cm]{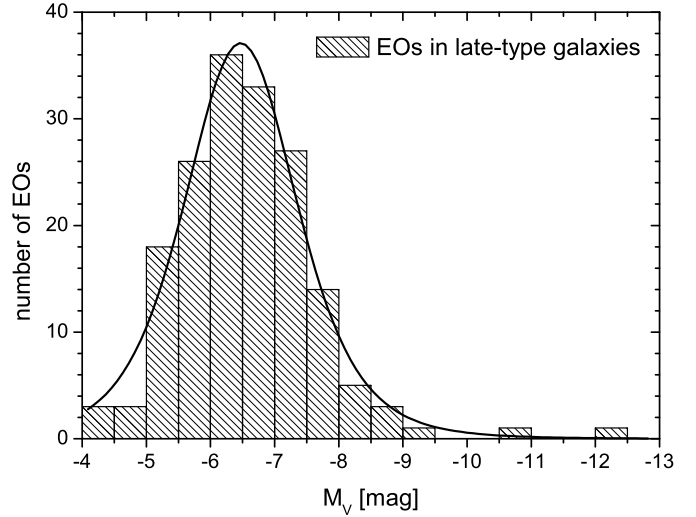}
\caption{Histogram of the number of EOs in late-type galaxies at different total V-magnitudes. 
The black line represents a fitted t$_5$ luminosity function.}
\label{fig_eolumilate}
\end{figure}

\begin{figure}[t]
\centering
\includegraphics[width=8.8cm]{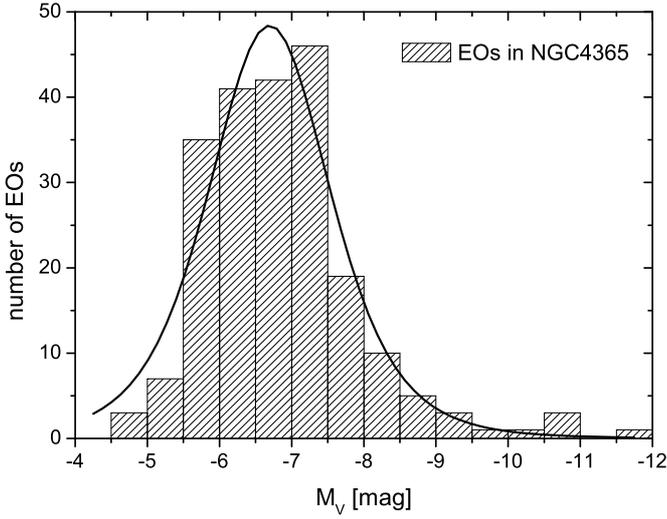}
\caption{Histogram of the number of EOs in the elliptical galaxy NGC4365 at different 
total V-magnitudes. The black line represents a fitted t$_5$ luminosity function.}
\label{fig_eolumiN4365}
\end{figure}

\begin{figure}[t]
\centering
\includegraphics[width=8.8cm]{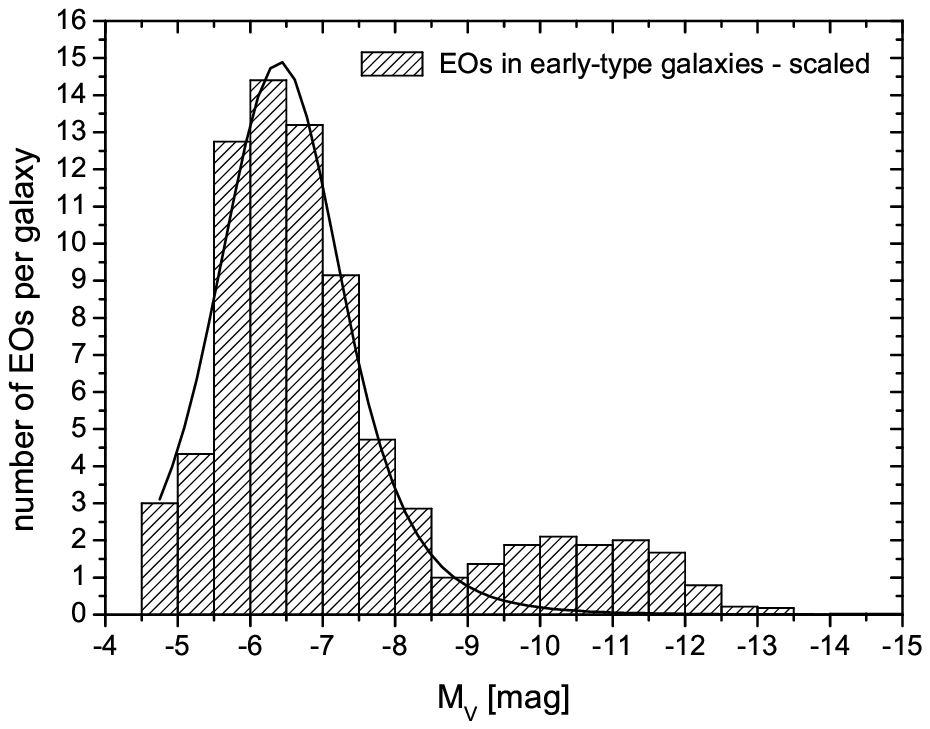}
\caption{Histogram of the normalized number of EOs in early-type galaxies at different 
total V-magnitudes scaled by the number of galaxies having observations at the individual bins. 
The black line represents a fitted t$_5$ luminosity function.}
\label{fig_eolumiearly}
\end{figure}

\begin{figure}[t]
\centering
\includegraphics[width=8.8cm]{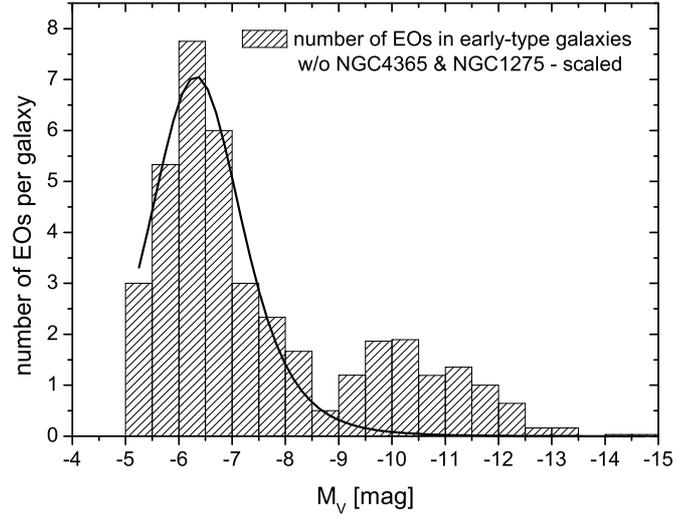}
\caption{Histogram of the normalized number of EOs in early-type galaxies except for the large
EO samples of NGC4365 and NGC1275 at different total V-magnitudes scaled by the number of galaxies 
having observations at the individual bins. The black line represents a fitted t$_5$ luminosity function.}
\label{fig_eolumiearlywo4365a1275}
\end{figure}

\subsection{EO Luminosity Functions} 
A common way of parameterization of samples of astronomical objects are luminosity functions. 
\cite{secker} analyzed compact GCs in the Milky Way and M31 and concluded that their luminosity 
functions are well represented by so-called Student t$_5$ functions 
\begin{eqnarray} 
LF \propto \left[1+\frac{(M_{\mathrm{V}}-M_{\mathrm{V,TO}})^2}{5\sigma_{\mathrm{V}}^2} \right]^{-3},\label{lumfunction}
\end{eqnarray}
where $M_{\rm V}$ are absolute V-band magnitudes of star clusters, $M_{\rm V,TO}$ is the turnover of the 
luminosity function and $\sigma_{\rm V}$ is the dispersion of the t$_5$ function. For compact GCs, the 
turnover of the luminosity function, $M_{\rm V,TO}$, is very constant for various types of galaxies, 
making it a reasonable distance estimator \citep[see][and references therein]{rejkuba12}. The mean 
turnover luminosity for the Milky Way and 18 nearby galaxies is $M_{\rm V,TO}$ = $-$7.66$\pm$0.09 mag. 

\subsubsection{EOs in Late-Type Galaxies}
A histogram of the number of EOs at different total luminosities of late-type galaxies is shown 
in Fig. \ref{fig_eolumilate}. A fit of the luminosity function according to Eq. \ref{lumfunction}
is added. The turnover of the luminosity function is $M_{\rm V,TO}$ = $-$6.47$\pm$0.03 
mag and the dispersion of the t$_5$ function is $\sigma_{\rm V}$ = 0.91$\pm$0.03 mag. 
The peak of the EO luminosity function is about one magnitude fainter than the typical peak of the GC luminosity function. 

For the histogram in Fig. \ref{fig_eolumilate} all 171 EOs associated with 
late-type galaxies are used. As discussed in Section \ref{sect_confirmed}, only a fraction of 
EOs were confirmed by follow-on
spectroscopy. Consequently, it cannot be excluded that some background galaxies modified the exact
result of the luminosity function. The number of confirmed EOs in late-type galaxies is 43. 
The mean total luminosity of the 43 EOs is M$_{\rm V}$ = $-$6.40 mag. 
Considering the low number of objects, this value
is quite close to the fitted value for all candidate EOs, indicating that the EOs in late-type 
galaxies indeed have a fainter peak of the luminosity function than compact GCs.

Considering the very low surface brightness of faint and extended EOs (see 
Sect.~\ref{sect_confirmed}), a fair fraction of very extended and faint EOs is most likely below 
the detection limit of extra-galactic surveys. The true turnover of the EO luminosity function 
might therefore be at even lower luminosities.

\subsubsection{EOs in Early-Type Galaxies}\label{sectlumiearly}
Figure \ref{fig_histmvall} shows that EOs in early-type galaxies show a bimodal distribution which 
peaks at about $M_{\rm V}= -6.5$ mag and $-$11.0 mag and has a clear minimum between $-$8.5 and $-$9 mag.
On the other hand, Figure \ref{fig_dgal}b demonstrates that for a large fraction of early-type
galaxies only high-luminosity objects were considered. This is partly due to detection limits 
especially at large distances, but also due to the fact that since the discovery of UCDs in the 
Fornax Cluster by \cite{hilker99} and \cite{drinkwater00}, much effort has been made to detect 
and to analyze EOs brighter than about M$_{\rm V}$ = $-$10 mag, while fainter EOs were neither in the focus 
of UCD studies nor in those investigating GCs.

A number of GC surveys applied size limits to reduce the contamination of background 
galaxies. For instance, the GC surveys covering 100 galaxies of the Virgo Cluster \citep{jordan05} and 
43 galaxies of the Fornax Cluster \citep{masters10} applied a size limit of $r_{\rm eff} <$ 10 pc to 
reduce the contamination by background galaxies. As a side effect, they excluded also all EOs from 
their GC catalogs.

The very different approaches for objects with low and high luminosities have a significant influence
on the luminosity function. One example is the galaxy M85 of the Virgo Cluster. 
Four EOs are found with luminosities brighter than M$_{\rm V}$ = $-$8.5 mag \citep{hasegan,chies11}. 
\cite{peng} used the same Virgo Cluster survey data as \cite{jordan05} to search for diffuse star clusters 
and concluded that the galaxy M85 has about 30 EOs with luminosities between M$_{\rm V}$ = $-$5.5 and $-$8.5 mag. 
While the low luminosity objects would dominate for this galaxy, our catalog contains only the four bright 
M85 EOs of \cite{hasegan} and \cite{chies11} as \cite{peng} have not published a catalog of their EOs.
In the same field of view \cite{jordan05} found 211 compact GCs in M85. For this specific galaxy, the 
EOs would add about 15\% to the GC sample. 

Another example is the giant elliptical galaxy NGC\,4365, which was covered by eight HST ACS fields 
\citep{blom}. Seven HST ACS fields provide a very good coverage of the inner 35 kpc of the galaxy 
while the last field pointed to the halo delivering clusters with projected distances between about 
40 and 60 kpc. \cite{blom} found in total 2038 GC and 217 EO candidates. 
For this galaxy, EOs add about 10 percent to the GC sample. 
Only 5 of the 217 EOs are brighter than M$_{\rm V}$ = $-$10 mag, 19 EOs have a luminosity between 
M$_{\rm V}$ = $-$8 and $-$10 mag, and 148 EOs have one between M$_{\rm V}$ = $-$6 and $-$8 mag. Again, the number of
low-luminosity EOs is significantly larger than the number of bright EOs. 
The EO candidates of NGC4365 form the largest EO sample of an elliptical galaxy. 
Figure~\ref{fig_eolumiN4365} shows the histogram of the number of EO candidates of NGC4365 as a function
of luminosity. The black line shows the corresponding luminosity function having a turnover luminosity
of $M_{\rm V,TO}$ = $-$6.69$\pm$0.07 and a dispersion of the t$_5$ function of $\sigma_{\rm V}$ = 0.88$\pm$0.07 mag.
The turnover luminosity is at slightly brighter luminosities as for the late-type galaxies.

In their central HST field, \cite{blom} found 681 compact clusters and 30 EOs, i.e. the EOs add 
about 4.4\% to the GC sample in the central region of NGC\,4365. For the 100 Virgo 
galaxies \cite{jordan05} found in single HST fields centered on the individual galaxies in total 12763 
compact GCs. An EO fraction of 5\% would yield for these 100 galaxies an EO population of 638 EOs, which 
is about ten times the number of high luminosity objects in our catalog associated with these Virgo Cluster 
galaxies. 

Consequently, an interpretation of the bimodal luminosity distribution needs to take into account the 
varying number of galaxies building the sample at each luminosity bin. Only seven early-type galaxies 
have observations of EOs fainter than M$_{\rm V}$ = $-$8.5 mag and only four galaxies have observations of EOs 
fainter than M$_{\rm V}$ = $-$6.5 mag. For EO luminosities from M$_{\rm V}$ = $-$8.5 mag to $-$10.5 mag, the number of 
observed galaxies increases from 9 to 20. EOs brighter than M$_{\rm V}$ = $-$10.5 mag were 
observed in 33 galaxies. 

\begin{figure}[t]
\centering
\includegraphics[width=8.7cm]{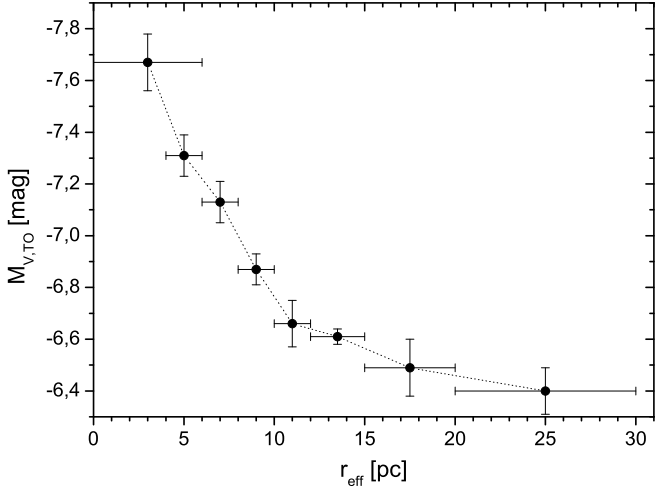}
\caption{Turnover of the luminosity function, $M_{\rm V,TO}$, of GC and EOs plotted as a function 
of the effective radius. The error bars in $r_{\rm eff}$ indicate the individual bin-sizes, the error bars
in $M_{\rm V,TO}$ show the statistical error of the fit of the luminosity function to the data.}\label{fig_variolumi}
\end{figure}

Figure \ref{fig_eolumiearly} takes the varying number of galaxies into account, i.e. the number of EOs in
each bin is divided by the number of galaxies contributing to this luminosity bin. 
The second peak at high luminosities has decreased considerably, but it has not entirely 
vanished, demonstrating that the large number of high-luminosity EOs associated with the most luminous 
galaxies is not a simple size-of-sample effect.
A fit of the luminosity function according to Eq. \ref{lumfunction}
is added to Fig. \ref{fig_eolumiearly}. The peak of the luminosity function is at 
$M_{\rm V,TO}$ = $-$6.40$\pm$0.06 mag and the dispersion of the t$_5$ function is 
$\sigma_{\rm V}$ = 0.89$\pm$0.07 mag. 
The values change only slightly to $M_{\rm V,TO}$ = $-$6.40$\pm$0.05 mag 
and $\sigma_{\rm V}$ = 0.87$\pm$0.06 mag when
the high-luminosity tail (i.e. objects brighter than $M_{\rm V}$ = $-$9 mag) is excluded from the fit.

The EO sample of early-type galaxies is dominated by the two galaxies NGC4365 and NGC1275, which
have 217 and 84 EO candidates, respectively. In order to verify that the results for early-type
galaxies are not biased towards these two galaxies, we have repeated the exercise of scaling the
number of objects per luminosity bin by the number of relevant galaxies excluding NGC4365 and NGC1275
from the sample. Figure~\ref{fig_eolumiearlywo4365a1275} shows the resulting histogram and a fitted
luminosity function having $M_{\rm V,TO}$ = $-$6.32$\pm$0.10 mag and $\sigma_{\rm V}$ = 0.89$\pm$0.12 mag. 
The results of the sample without NGC4365 and NGC1275 agree well with the results of the entire EO 
sample of early-type galaxies.

The results for early-type galaxies are very similar to the results from late-type galaxies.
The main difference between the luminosity functions is the tail of high-luminosity objects 
associated with early-type galaxies. 

\citet{mieske12} studied a sample of confirmed GCs and UCDs to adress the question
whether there is an over-population of UCDs with respect to compact GCs. They concluded that 
the number of UCDs is consistent with a continuation of the GC luminosity function towards 
bright magnitudes. In this paper, we demonstrate that there is an over-population of bright 
EOs if compared with the EO luminosity function. 
While the \citet{mieske12} sample is well defined in the sense that all objects were 
spectroscopically confirmed, most of them have no measured size. They used all UCDs irrespective 
of their size and compared the number of UCDs with the GC luminosity function which
has a turnover luminosity that is about one magnitude brighter than the EO luminosity function. 
In additon, the number of GCs is about 10 times larger than the number of EOs. 
Consequently, the results of \citet{mieske12} cannot easily be compared with our results as
the samples are largely independend from each other. Larger EO and GC samples of a number of
early type galaxies covering the entire luminosity range from $M_{\rm V}$ = $-$4 to $-$14 mag 
are needed to answer the question whether there is a general overpopulation of bright 
EOs in early-type galaxies or whether it is a specific feature seen only in special environments.

\subsubsection{Trends of the turnover luminosity with effective radius}
In the previous sections, we concluded that EOs in early and late-type galaxies have basically the
same turnover of the luminosity function, which is about one magnitude fainter than that of compact GCs. 

In this section, we explore the trend of the turnover luminosity with increasing effective radii.
In order to increase the number of objects per $r_{\rm eff}$-bin, we combine the data for GCs and EOs of
early and late-type galaxies. We exclude the brightest objects in the UCD regime to avoid the second peak
as seen in Figs. \ref{fig_histmvall} and \ref{fig_eolumiearly} and focus on objects fainter than M$_{\rm V}$ = $-$9 mag.

For the compact GCs with effective radii below 6 pc, we derive a turnover of the luminosity function of
$M_{\rm V,TO}$ = $-$7.67$\pm$0.11 mag, which is in very good agreement with the mean turnover luminosity 
for the Milky Way and 18 nearby galaxies, $M_{\rm V,TO}$ = $-$7.66$\pm$0.09 mag \citep{rejkuba12}.

Figure \ref{fig_variolumi} shows the trend of the turnover luminosity as a function of the effective radius.
The turnover luminosity decreases continuously from $M_{\rm V,TO}$ = $-$7.67 mag to 
$M_{\rm V,TO}$ = $-$6.66 mag
at the $r_{\rm eff}$-bin between 10 and 12 pc. For larger effective radii, the turnover luminosity decreases
considerably slower to values of $M_{\rm V,TO}$ = $-$6.40 mag at the $r_{\rm eff}$-bin between 20 and 30 pc.

On the basis of the available data, we conclude that the turnover of the luminosity function depends 
significantly on the effective radii of star clusters and that the slope of the varying $M_{\rm V,TO}$ 
is steeper for GCs than for EOs.

However, the edge in the luminosity function at $M_{\rm V}$ = $-$5.0 mag for late-type 
and at $M_{\rm V}$ = $-$5.5 mag for early-type galaxies indicates that the samples are fairly 
incomplete at very low luminosities. The true turnover of the EO luminosity function 
is therefore expected to be at even lower luminosities, which might lead to a steeper slope 
in the EO regime. Considerably larger and more complete data-sets especially at low-luminosities, 
consisting of confirmed star clusters, are necessary to confirm the trend of the luminosity functions 
from GCs to EOs as shown in Fig. \ref{fig_variolumi} and to substantiate the idea that this trend 
is a general feature of GCs and EOs in early- and late-type galaxies. 

\subsection{Spatial Distribution of EOs} \label{incomplete_coverage}
In the Milky Way, the 11 EOs have galacto-centric distances between 16 and 125 kpc with a median
distance of 72 kpc. In contrast, the GCs of the Milky Way are at considerably lower distances. 
The median distance of the GCs is 4.8 kpc. 
For M31 five out of 20 EOs have projected distances smaller than 10 kpc and 11 EOs have projected distances
larger than 20 kpc \citep{peacock,huxor08}. The two EOs associated with M33 have projected distances of 12.5 
and 28.4 kpc \citep{stonkute,huxor09}.

The most serious constraint of EO catalogs of galaxies outside the Local Group is the very limited 
field of view of the HST ACS instrument, as EOs at large distances from the Galaxy can only be spatially 
resolved by HST. The field of view of the ACS instrument is 202\arcsec\ and the pixel size is 0.05\arcsec. 

The dotted curve in Fig.~\ref{fig_dgal}a indicates the largest possible projected radius of a single 
HST ACS image centered on a galaxy at a given distance. It visualizes the area covered by one Hubble field. 
At small distances a couple of HST ACS fields are necessary to scan a galaxy for EOs, 
whereas at large distances the galaxy and part of the halo are completely covered by one HST ACS field.

\begin{figure}[t]
\centering
\includegraphics[width=8.6cm]{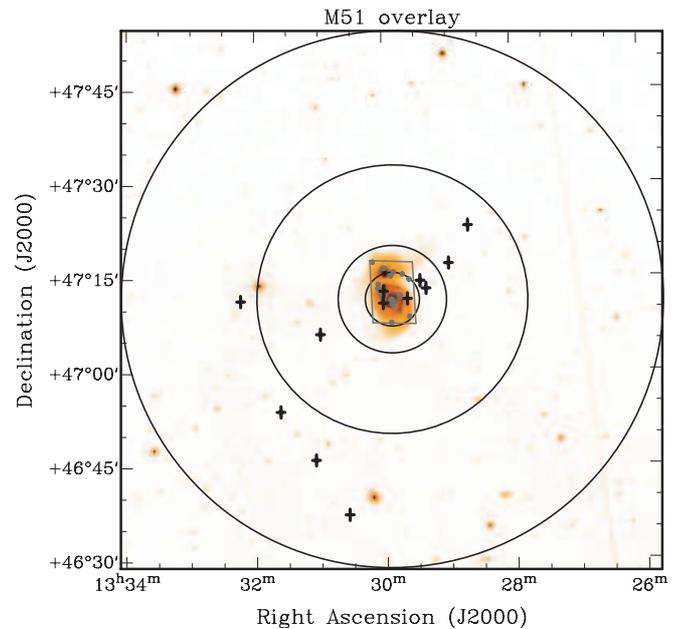}
\caption{M51 EOs (grey dots) overlayed on a Digitized Sky Survey 2 (DSS2) image of M51. 
The grey lines indicate the size of the HST mosaic of 6 ACS images used by \cite{hwang08}
to find EOs in M51. The black circles indicate projected distances from M51 of 10, 20, 50, and 100 kpc. 
For comparison, the projected position of the Milky Way EOs, if the Milky Way would be seen face-on 
at the distance and the position of M51, are added as black crosses.}\label{fig_m51}
\end{figure}

At the distance of the Whirlpool Galaxy M51 of about 8 Mpc, the field of view of 202\arcsec\ 
and the pixel size of 0.05\arcsec\ correspond to 7.8 kpc and 2 pc, respectively.
While the pixel size is well suited to resolve EOs, a number of ACS images are needed to cover the
entire galaxy. Figure \ref{fig_m51} shows an image of the Digitized Sky Survey 2 
(DSS2\footnote{The Digitized Sky Survey data used in this paper have been taken from the ESO Archive, 
see http://archive.eso.org/dss/dss}) of M51 and the area covered by the mosaic of six ACS images 
used by \cite{hwang08} to find EOs in M51, demonstrating that even six HST ACS images do not cover 
the entire stellar body of the interacting galaxy pair. 

The HST mosaic of M51 covers an area of 15.7 kpc $\times$ 23.6 kpc. This coverage is 
well suited to detect EOs related to the stellar bodies of the two galaxies and EOs in the 
lower halo.
While halo EOs at considerably larger distances from their host galaxies might 
be found by chance in projection to the main body of the host galaxy, the probability is relatively low. 
Considering a line-of-sight of $\pm$100 kpc, the volume covered by the HST mosaic is 
15.7 $\times$ 23.6 $\times$ 200 kpc$^3$, or 7.4 $10^{4}$ kpc$^3$. In contrast, the volume with a radius of 
100 kpc, which would enclose the EOs, is $\frac{4}{3}\pi$100$^3$ kpc$^3$ or 4.2 $10^{6}$ kpc$^3$. 
The HST mosaic of six ACS images covers less than 2 percent of the volume expected to contain EOs.

Figure \ref{fig_m51} shows for comparison also the location of the Milky Way EOs, if the Milky Way would 
be seen face-on at the distance and the position of M51. The six most distant EOs of the Milky Way 
have large galacto-centric distances between 70 and 125 kpc, or projected 
distances between 41 and 84 kpc in Fig. \ref{fig_m51}. This figure demonstrates that only 3 of 11 EOs 
would  be located within the HST mosaic. Consequently, also a number of EOs of M51 are expected to have 
considerably larger projected distances beyond the currently covered survey area. 

In addition to EOs located in galactic halos, \cite{larbro00} and \cite{bro02} have discovered a 
population of EOs co-rotating with the disk of the lenticular galaxy NGC\,1023. These so-called faint 
fuzzies have similar structural parameters as halo EOs and are therefore not easily distinguishable 
from halo EOs projected onto the disk on the basis of imaging data alone. A fair fraction of EOs found 
in extragalactic surveys - especially those covering only the disk regions like the M51 survey \citep{hwang08}
- might therefore be associated with the disks and not the halos of these galaxies.

\begin{figure}[t!]
\centering
\includegraphics[width=8.8cm]{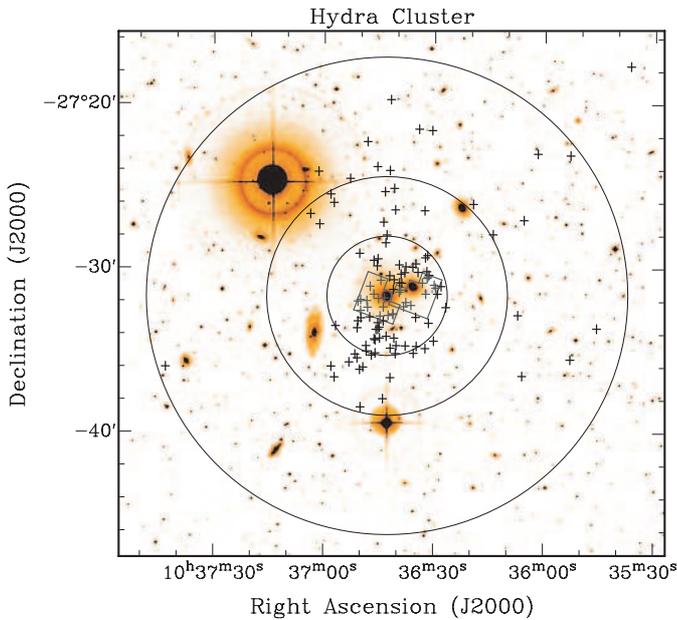}
\caption{Spectroscopically confirmed star clusters (crosses) of the Hydra Cluster \citep{misgeld11} 
overlaid on a DSS2 image of the Hydra Cluster. Only those clusters which are located within the two 
HST WFPC2 fields (marked by grey lines) have measured sizes. The black circles indicate 
projected distances from the central galaxy NGC3311 of 50, 100, and 200 kpc. }
\label{hydra}
\end{figure}

The Hydra Cluster is located at a distance of about 47.2 Mpc. \cite{misgeld11} searched with ground
based telescopes for massive star clusters in Hydra and detected and spectroscopically confirmed
118 objects with total V-band luminosities between $M_{\rm V}$ = $-$9.7 and $-$13.3 mag and projected
distances between 3 and 300 kpc. The median projected distance is 44 kpc.
Only 26 of the 118 stellar objects are located within two HST WFPC2 fields (see Fig. \ref{hydra}). 
19 of the 26 objects are EOs and the remaining 7 objects have effective radii between 8 and 10 pc. It is 
expected that also a large fraction of the remaining 92 objects in the outer halo are EOs. In addition, 
Fig. 7 of \cite{misgeld11} demonstrates that star clusters are not uniformly 
distributed in the halo. While some halo fields contain several UCDs, other fields have no UCDs at all.
Consequently, random samples of small fields in a halo cannot be used to extrapolate to the
entire population.

Another example is the giant elliptical galaxy NGC\,4365, which was covered by eight HST ACS fields 
\citep{blom}. Seven HST ACS fields provide a very good coverage of the inner 35 kpc of the galaxy. 
In their central HST field, \cite{blom} found 681 GC and 30 EO candidates, 
while the six HST fields surrounding the central field have on average 216 GCs and 26 EOs. 
The results demonstrate that the number density of compact clusters decreases rapidly towards larger 
projected distances while the number of EOs per HST field is almost constant. In addition, \cite{blom} 
searched for star clusters outside the HST fields with ground based telescopes and concluded that the 
cluster system extends out to projected radii of about 135 kpc and contains about three times the number 
of clusters as the HST fields. Consequently, there are several hundreds of EOs expected in the halo 
of NGC\,4365 outside the HST fields.

The three examples M51, NGC4365 and the Hydra Cluster are typical for the entire sample of galaxies with 
EOs. The vast majority of the studied galaxies have a reasonable coverage of the main stellar body, but a 
very poor coverage of the halo by HST observations. Only very distant galaxies like the elliptical 
galaxy ESO 325-G004 in the galaxy cluster Abell S0740 at a Galactic distance of 143 Mpc are well covered 
(including the halo) by one HST ACS field. However, the pixel size of the HST ACS of 0.05\arcsec\ corresponds 
to a linear scale of about 35 pc at this distance, which is too coarse to resolve rather compact EOs.

A considerably larger fraction of the halos of early and late-type galaxies needs to be covered
by HST observations to allow for a conclusive view on the spatial distribution of EOs and possible 
differences between early and late-type galaxies.

\subsection{An EO Formation Scenario}
\cite{phillipps}  interpreted the high-luminosity EOs as a new type 
of galaxy and reflected this 
interpretation in the name ``ultra-compact dwarf galaxy'' (UCD). \citet{bekki01,bekki03} suggested 
that UCDs are the remnants of dwarf galaxies which lost their dark matter halo and 
all stars except for their nucleus. 
Next to the interpretation as a galaxy, UCDs were also considered as high-mass versions 
of normal GCs \citep{mieske02}, or as merged massive complexes of star clusters 
\citep{krou98,fellhauer02a,bruens11}. In this paper, we will focus on the latter, i.e. 
the star cluster origin where EOs are the end products of merged star clusters.

A few decades ago `young massive star clusters' (YMCs) were found with globular cluster-like
properties. YMCs are found in all types of gas-rich galaxies and constitute a common class of 
star clusters. The definition of YMCs is rather author dependent. We adopt the definition of 
\cite{whitmore03} who defined `young' as having an age less than 500 Myr and `massive' as having 
masses ranging from $10^3\,M_\odot$ to $10^{8}\,M_\odot$. 
Individual YMCs were analyzed in detail by \citet{bastian06b}, \citet{mengel08}, and 
\citet{bastian09}. The combined data-set of the three publications demonstrates that the median 
size of YMCs in the mass range between $10^5\,M_\odot$ and $10^{6}\,M_\odot$ is about 4 pc. 
YMCs with masses of a few times $10^7\,M_\odot$ have only been observed in strong starburst 
environments like for example in the interacting galaxy NGC\,6745 \citep{de_grijs03} and the 
late-stage merger galaxies NGC\,7252 \citep{Maraston04} and Arp\,220 \citep{wilson06}.

Observations have shown that YMCs are often not isolated, but are part of larger structures called CCs 
\citep[e.g.][]{bastian06}. The CCs contain few to hundreds of YMCs spanning up to a few hundred parsecs in 
diameter. The mass of a CC is the sum of its YMC constituents. Little is known about the detailed distribution 
of the individual star clusters inside a CC and about their velocity distribution, largely because the existence 
of CCs had not been realized fully until only very recently.
However, the observations show that most CCs have a massive concentration of star clusters in their centers 
and a couple of isolated star clusters in their vicinity.

\begin{figure}[t]
\centering
\includegraphics[width=8.7cm]{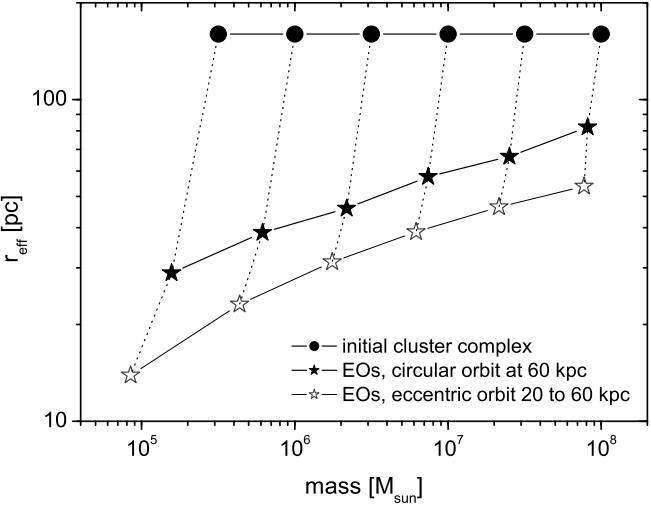}
\caption{Diagram of $r_{\rm eff}$ as a function of mass for the modeled merger objects on a 
circular orbit at 60 kpc (black stars) and on an eccentric orbit between 20 and 60 kpc (open stars) 
after a dynamical evolution of 5 Gyr \citep[see][for details]{bruens11}. For comparison, the parameters 
of the initial CCs (black circles) are included. The initial CCs are connected with the corresponding merger
objects by dotted lines.}
\label{fig_Sims}
\end{figure}

\begin{figure}[t]
\centering
\includegraphics[width=8.7cm]{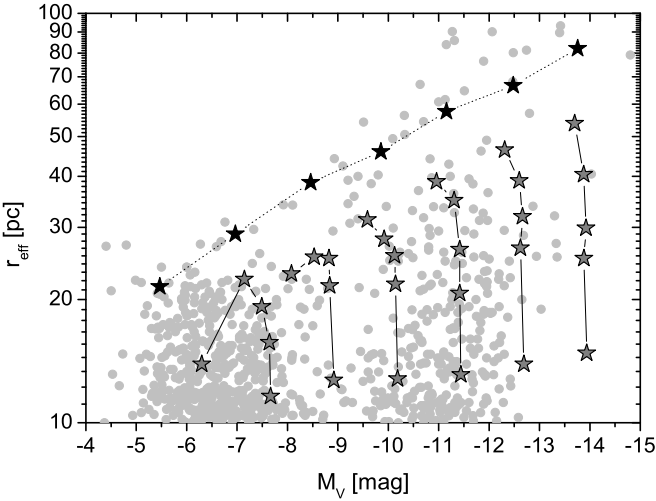}
\caption{Diagram of $r_{\rm eff}$ as a function of the total luminosity 
of the observed EOs (grey circles) and the modeled merger objects (stars) from \cite{bruens11}. 
Merger objects on eccentric orbits between 20 and 60 kpc are plotted in dark grey. Objects with the same 
initial CC mass but different initial CC effective radii between 10 and 160 pc are connected by black lines. 
Merger objects on a circular orbit at 60 kpc and an initial effective radius of the CC of 160 pc are plotted 
as black stars connected by dotted lines.}
\label{EOandSims}
\end{figure}

Examples for CCs are the knots of the interacting galaxies NGC\,4038 and NGC\,4039 at a distance of about 
20 Mpc,
aka the Antennae, which have typical masses of the order of $M_{\rm knot}$ = $10^6$ to $10^7\,M_\odot$ 
and sizes of a few hundred parsecs \citep{whitmore05}. The knots typically consist of about 
30 young massive star clusters with masses greater than $10^4\,M_\odot$ and about 60 lower-mass 
clusters \citep{whitmore10}. 
\citet{whitmore05} found that the cluster to cluster velocity dispersion in the knots is 
small enough to keep them gravitationally bound leading to merging 
of clusters in the central region of the knots. 
The collision between the two gas-rich 
Antennae galaxies triggered the formation of super giant molecular complexes with masses up to 
$9\cdot10^8\,M_\odot$ \citep{wilson03}. It is expected that these super massive gas clouds 
will be the birth sites of very massive CCs. Some starburst galaxies like for example Arp 220 at a 
distance of 77 Mpc host YMCs/CCs as massive as $10^7\,M_\odot$ with ages less than 10 Myr \citep{wilson06}. 
Arp 220 also represents two colliding spiral galaxies, but already in the end phase of the merging process.
The late-stage merger galaxy NGC~7252 hosts a very massive star cluster (W3) with an age between 300 
and 500 Myr. It has a mass of about $8 \cdot 10^7$ M$_{\sun}$ and an effective radius of 
$R_{\rm eff}$ = 17.5 pc \citep{Maraston04}. From its structural parameters W3 may be classified 
as a young version of a UCD. The young age of W3 precludes an origin as a remnant nucleus of a stripped 
dwarf galaxy. W3 may instead have evolved from a CC by merging of its constituent star clusters 
\citep{fellhauer05}. 
Another example is the Tadpole galaxy (UGC 10214), which is a disrupted barred spiral galaxy at a 
distance of about $130$~Mpc. The galaxy shows a long tidal tail of stars, which hosts also some CCs.
The most luminous and largest CC in the tail has a mass of the order $M_{\rm CC}$ = $10^6$ M$_{\sun}$, 
an effective radius of  $r_{\rm eff}$ = 160 pc, a cut-off radius of about $r_{\rm cut}$ = 750 pc and it
is located at approximately 60 kpc from the center of the galaxy \citep{tran}.

Since galaxy-galaxy mergers are anticipated to have been more common during early cosmological times 
it is expected that star formation in CCs has been a significant star formation mode during this epoch. 
Indeed, the preponderance of clumpy galaxies \citep[and references therein]{elmegreen07} indicates
that early gas-rich galaxies went through an epoch of profuse CC formation. 
\cite{bournaud08a} performed high-resolution modeling of a galaxy interaction that lead to a
merger remnant comparable to an elliptical galaxy. In these models, super star clusters with masses
of a few $10^7$ M$_{\sun}$ and sizes up to about 150 pc formed in the halo of this merger remnant. 
\cite{bournaud08b} demonstrated that these star clusters are gravitationally stable.

The dynamical evolution of CCs with a large range of initial configurations and on various orbits
was studied in a number of publications 
\citep[e.g.][]{krou98,fellhauer02a,fellhauer02b,fellhauer05,bekki04,bruens09,bruens10,bruens11}. These studies 
demonstrated that the merging star cluster scenario provides a mechanism for the formation of EOs.

In a previous paper \citep{bruens11}, we varied sizes and masses of the modeled CCs covering a matrix of 
5x6 values with effective radii between 10 - 160 pc and CC masses between $10^{5.5}$ - $10^{8}$ M$_{\sun}$.
These models were placed on eccentric orbits between 20 and 60 kpc and were simulated over a period of
5 Gyr\footnote{\cite{bruens10} demonstrated that the structural parameters of the merger objects change
only marginally after the merging process is concluded after a few Gyr.}. 
In addition, models with CCs having effective radii of 160 pc were placed on a circular orbit at
a distance of 60 kpc. Figure \ref{fig_Sims} shows the location of the initial CCs having $r_{\rm eff}$ = 160 pc
and the corresponding modeled merger objects on the eccentric and the circular orbit after 5 Gyr of evolution. 
Due to the merging process and the external tidal field, the final masses and effective radii of the merger 
objects are considerably lower than those of the initial CCs. The comparison between the eccentric and the 
circular orbits demonstrates that the tidal field plays a crucial role for the final parameters of the 
merger objects.

Figure \ref{EOandSims} shows effective radii and estimated total V-band luminosities 
of the models of \cite{bruens11} using a mass-to-light ratio of 3 and the 
parameters of the 813 EOs as given in Fig. \ref{fig_reffmv}. Figure \ref{EOandSims} demonstrates 
that the observed and simulated EOs cover the same parameter space and show the same trend of 
increasing effective radii with increasing luminosity.

The merged cluster scenario also helps to understand the difference between the mass function of
YMCs, which is a power-law with a slope of $-2$ and the GC/EO luminosity functions, which have 
a bell-shaped function as given in Eq. \ref{lumfunction}. Most publications focused on the
destruction of low-mass clusters \citep[for a recent review see][and references therein]{Portegies}. The merged
cluster scenario predicts that dozens of low-mass YMCs merge to form one EO. As EOs add
approximately 10\% to the number of GCs, it is expected that the total number of low-mass clusters 
needed to build up the EOs is larger than the actual number of remaining GCs.
Consequently, the merged cluster scenario predicts that a fair fraction of the low-mass YMCs evolve 
via merging into EOs modifying the original power-law luminosity function.

\begin{figure}[t!]
\centering
\includegraphics[width=8.7cm]{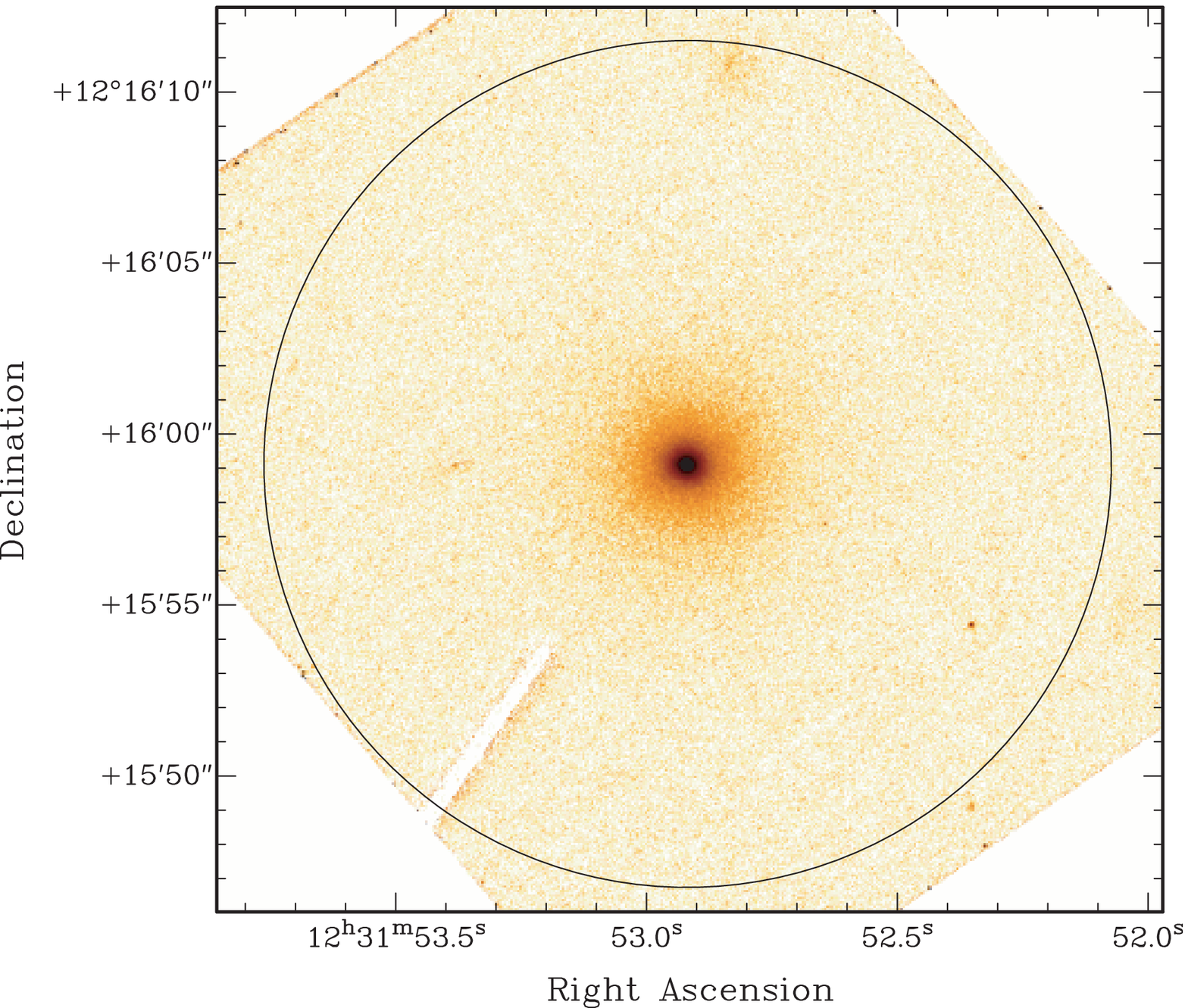}
\caption{High resolution HST ACS image of VUCD7 in the F606W filter. The black circle has a projected 
radius of 1 kpc indicating the extent of the surface brightness profile shown in Figure \ref{VUCD7profile}.}
\label{VUCD7image}
\centering
\includegraphics[width=8.7cm]{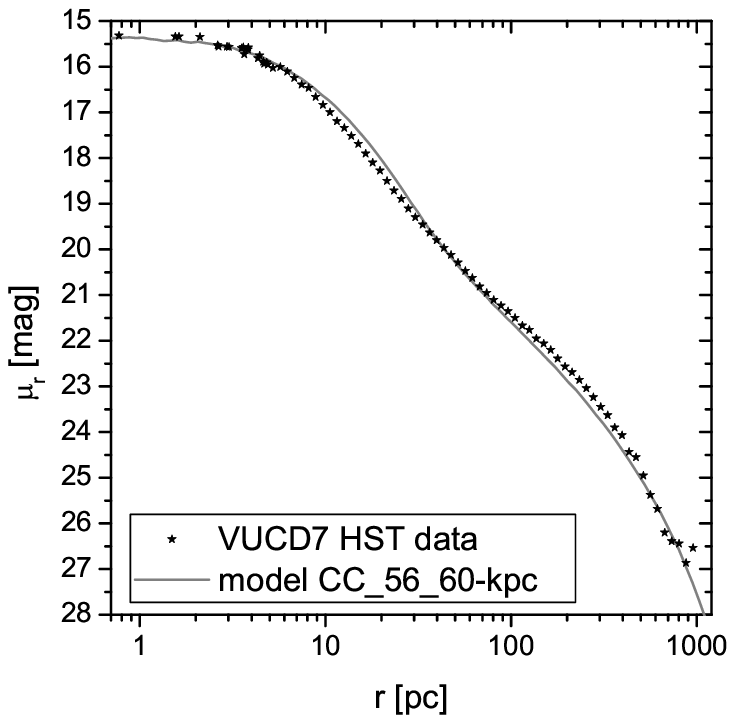}
\caption{Surface brightness profile compiled from archival HST ACS data (black stars) of the EO VUCD7 and 
the surface brightness profile of model CC\_56\_60-kpc (grey line) from \cite{bruens11} using a 
mass-to-light ratio of 2.8. The model and the observed EO show a very similar two-component or 
"core-halo" structure.}
\label{VUCD7profile}
\end{figure}

YMCs and CCs with masses below $M$ = $10^6$ M$_{\sun}$ have been observed in all types of galaxies containing
sufficient amounts of gas to form star clusters, while high-mass YMCs and CCs ($M$ $\ga $ few$\times 10^6$ M$_{\sun}$) 
were so far only observed in major merger events like the Antennae, Arp 220, or NGC\,7252, which may be in 
the process of becoming elliptical galaxies.
This explains the different EO luminosity functions for late-type and early-type galaxies: only the
major interactions that create the early-type galaxies build CCs massive enough to form the EOs considerably
brighter than about $M_{\rm V}$ = $-$10 mag.

The evolution of low-mass EOs in a weak gravitational environment has also been studied by \cite{hurley}, 
who performed direct N-body models of extended low-mass star clusters incorporating a stellar mass 
function and stellar evolution. Internal evolution processes of the star clusters lead to considerably 
larger effective radii compared to the initial values. They concluded that extended star clusters with an 
initial mass of 5.8 $10^{4}$ M$_{\sun}$ are sufficiently stable to survive a Hubble-time in a weak 
gravitational field environment.

\cite{murray} addressed the question why compact star clusters are absent in the high-mass regime,
i.e. at masses larger than few $10^{6}$ M$_{\sun}$. \cite{murray} modelled proto-clusters and found that
high-mass clusters become optical thick to infra-red radiation leading to a modified initial mass function
of the cluster stars, to a larger mass-to-light ratio, and to considerably larger sizes of the proto-clusters. 
While the idea of optically thick proto-clusters explains the absence of compact clusters in the high-mass
regime, it cannot explain the existence of EOs over the large observed mass-range.

An alternative formation scenario for high-mass EOs is the galaxy threshing scenario, where nucleated 
dwarf galaxies loose, during a heavy interaction with a larger galaxy, almost all stars of the main body 
except for the nucleus \citep{bekki01}. This formation scenario connects EOs with compact elliptical 
galaxies like M32, as this rare species of compact galaxy is also expected to be the end-product of an interaction
that stripped the bigger part of a formerly larger galaxy. 
\cite{bekki03} performed numerical simulations to demonstrate that 
nucleated dwarf galaxies can indeed evolve into EOs if they are on a highly eccentric orbit. 
However, these studies on the threshing scenario focussed on the UCD regime. So far, it was
not demonstrated that the threshing scenario is able to reproduce also low-mass EOs.

According to \cite{bekki03}, a two-component "core-halo" surface brightness profile is a major 
prediction of the galaxy threshing scenario, where the core is the former nucleus and the halo is the
remnant of the main body of the former galaxy.

Three of the four most luminous confirmed EOs, i.e. VUCD7, UCD3, and M59cO, show a clear two-component or 
core-halo surface brightness profile. Following the results of \cite{bekki03}, this kind of structure was 
used as evidence for a galactic origin of UCDs \citep[e.g.][]{hasegan,chilingarianmamon11,norris11}.
While UCD3 is largely overlapping with a background spiral galaxy and M59cO has a projected distance to M59 
of only 9 kpc, the most extended and most luminous of the three objects, VUCD7, is located about 83 kpc from M87 
and has neither disturbing foreground nor background objects in its vicinity. Figure \ref{VUCD7image} shows
an HST ACS high-resolution image in the F605W band\footnote{The HST ACS image of VUCD7 was taken from the 
Hubble Legacy Archive, see http://hla.stsci.edu/hlaview.html}. The HST data
allow to compile a surface brightness profile out to large projected radii of about 1 kpc.

Figure \ref{VUCD7profile} shows the surface brightness profile of VUCD7 compiled from the archival HST ACS 
data presented in Fig. \ref{VUCD7image} using a median filter for the radial bins to exclude faint emission
from possible foreground or background objects. The observed profile of VUCD7 shows a clear two-component 
structure. Whereas most of the models from \cite{bruens11} show a single component profile, the 
surface brightness profiles of the most massive and most extended models on a circular 60 kpc orbit, which 
have a comparable mass and effective radius as VUCD7, show a two-component profile. 
Figure \ref{VUCD7profile} demonstrates that the surface brightness profile of model "CC\_56\_60-kpc" from
\cite{bruens11} shows a very simular two-component structure as the observed VUCD7. Consequently, a two-component 
"core-halo" surface brightness profile cannot be used as evidence for the stripping scenario as also the 
merged cluster scenario explains this specific structure of the most extreme UCDs.

In addition, \cite{bruens09} discussed the origin of the EOs, or faint fuzzies, associated
with the disk of NGC1023 in the context of merged cluster scenario and demonstrated that the observed 
structural parameters of the faint fuzzies are in excellent agreement with the modelled merger objects. 

In conclusion, the merged star cluster origin for EOs explains the existence of EOs over the entire 
observed range of luminosities of ECs and UCDs, the trends in the structural parameters, and the 
differences between early and late-type galaxies at the high-luminosity end.

\section{Summary \& Conclusions}\label{summary}
We searched the available literature to compile the largest possible catalog of star clusters 
with effective radii larger than 10 pc. As there is no clear distinction between ECs and UCDs, 
both types of objects are called extended stellar objects -- abbreviated "EOs" -- in this paper.

In total, we compiled a catalog of 813 EOs of which 171
were found associated with late-type galaxies and 642 EOs associated 
with early-type galaxies. The main results presented in this paper are

\begin{enumerate}
  \item EOs cover a luminosity range from about $M_{\rm V} = -4$ to $-$14 mag. However, 
  almost all EOs brighter than $M_{\rm V} = -10$ mag are associated with giant elliptical galaxies.
  \item At all values of $M_{\rm V}$ extended objects are found with effective radii between 10 pc and an upper 
  size limit, which shows a clear trend: the more luminous the object the larger is the upper size limit.
  This upper limit increases from about 30 pc at $M_{\rm V} = -5$ mag to about 100 pc at 
  $M_{\rm V} = -14$ mag. 
  \item The 175 confirmed EOs cover the same region in the $r_{\rm eff}$ vs. $M_{\rm V}$ space
  and show the same trends of increasing size with increasing luminosities as the 638 candidate EOs.
  \item For all luminosities, the majority of EOs have effective radii which are only slightly larger than 10 pc. 
  The median effective radius of EOs in late-type and early-type galaxies is 13.2 pc and 14.2 pc, respectively.
  \item The effective radii are increasing with increasing total luminosity of the host galaxy.
  \item For late-type galaxies there is no trend of the EO luminosities with the luminosity of the
  host galaxies, while for early-type galaxies there is a trend that the most luminous EOs are found
  associated with the most luminous galaxies.
  \item EOs and GCs form a coherent structure in the $r_{\rm eff}$ vs. $M_{\rm V}$ parameter space,
  which is well separated from the distribution of early type dwarf galaxies, except for the rare
  species of compact elliptical galaxies. Especially at the low-luminosity end considerably
  deeper observations are needed to answer the question, whether the prominent gap between ECs and
  dSph galaxies is real.
  \item For both EOs associated with early and late-type galaxies, the EO luminosity functions peaks 
  at about $-$6.5 mag, which is roughly one magnitude fainter than the turnover of the GC luminosity function. 
  The turnover luminosities decrease continuously between compact GCs and EOs for increasing effective radii.
  Considering the very low surface brightness of faint and extended EOs, a fair fraction of very 
  extended and faint EOs is most likely below the detection limit of extra-galactic surveys. The true 
  turnover of the EO luminosity function might therefore be at even lower luminosities.
\end{enumerate}

We discussed an EO formation scenario on the basis of a star cluster origin which explains the existence 
of EOs, the trends of the structural parameters and the differences between early and late-type 
galaxies over the entire range of luminosities of ECs and UCDs. In addition, a core-halo
surface brightness profile, as observed for some massive EOs like VUCD7, can be reproduced in the
merged CC formation scenario.

While this study presents the to day largest catalog of EOs, it suffers from substantial incompleteness,
predominantly with respect to the coverage of the galactic halos, but also with respect to varying
detection limits and the fairly low number of confirmed objects. 
Larger and more complete data-sets and additional information on parameters like metallicities,
mass-to-light ratios and age-estimates are necessary to draw final conclusions on the origin of EOs,
which in turn has the potential to shed light on the cosmologically important phase of galaxy formation.

\begin{acknowledgements}
      This work was supported by the German \emph{Deut\-sche For\-schungs\-ge\-mein\-schaft, DFG\/} 
      project number KR\,1635/29-1. We thank J.P. Madrid and S. Larsen for providing additional data on EOs
      not directly accessable from \cite{madrid11} and \cite{larbro00}. 
      We thank the anonymous referee for his helpful comments, which lead to a considerably improved paper.
\end{acknowledgements}

\end{document}